\documentclass[letterpaper]{article} % DO NOT CHANGE THIS
\usepackage{aaai25}  % DO NOT CHANGE THIS
\usepackage{times}  % DO NOT CHANGE THIS
\usepackage{helvet}  % DO NOT CHANGE THIS
\usepackage{courier}  % DO NOT CHANGE THIS
\usepackage[hyphens]{url}  % DO NOT CHANGE THIS
\usepackage{graphicx} % DO NOT CHANGE THIS
\usepackage{natbib}  % DO NOT CHANGE THIS AND DO NOT ADD ANY OPTIONS TO IT
\usepackage{caption} % DO NOT CHANGE THIS AND DO NOT ADD ANY OPTIONS TO IT
\usepackage{colortbl}
\usepackage{booktabs}       % professional-quality tables
\usepackage{amsfonts}       % blackboard math symbols
\usepackage{nicefrac}       % compact symbols for 1/2, etc.
\usepackage{microtype}      % microtypography
\usepackage{xcolor}         % colors
\usepackage{multirow}
\usepackage{enumitem}
\usepackage{comment}
\usepackage{amsmath}
\usepackage{subcaption}
\usepackage{soul}
\usepackage{hyperref}       % hyperlinks

\newcommand\gk[1]{{\color{red}{#1}}}

\newcommand{\squishlist}{
   \begin{list}{$\bullet$}
    { \setlength{\itemsep}{0pt}      \setlength{\parsep}{0pt}
      \setlength{\topsep}{3pt}       \setlength{\partopsep}{0pt}
      \setlength{\listparindent}{-2pt}
      \setlength{\itemindent}{-5pt}
      \setlength{\leftmargin}{1em} \setlength{\labelwidth}{0em}
      \setlength{\labelsep}{0.5em} } }
      
\newcommand{\squishend}{
    \end{list}  }

    \newcommand{\squishlistenum}{
   \begin{enumerate}
    { \setlength{\itemsep}{0pt}      \setlength{\parsep}{0pt}
      \setlength{\topsep}{3pt}       \setlength{\partopsep}{0pt}
      \setlength{\listparindent}{-2pt}
      \setlength{\itemindent}{-5pt}
      \setlength{\leftmargin}{1em} \setlength{\labelwidth}{0em}
      \setlength{\labelsep}{0.5em} } }

\newcommand{\squishendenum}{
    \end{enumerate}  }

\def \ie {{\it i.e.},~}

\def \eg {{\it e.g.},~}

\newcommand{\ben}{\begin{enumerate}}
\newcommand{\een}{\end{enumerate}}
\newcommand{\beq}{\begin{equation}}
\newcommand{\eeq}{\end{equation}}
\newcommand{\beqa}{\begin{eqnarray}}
\newcommand{\eeqa}{\end{eqnarray}}
\newcommand{\bit}{\begin{itemize}}
\newcommand{\eit}{\end{itemize}}
\newcommand{\btab}{\begin{tabular}}
\newcommand{\etab}{\end{tabular}}
\newcommand{\cond}{{\,|\,}}
\newcommand{\semic}{{\,;\,}}

\newcommand{\noprint}[1]{}

\newcommand{\mypara}[1]{\noindent\textbf{#1}}

\def \bfh {\mathbf{h}}

\def \calV {\mathcal{V}}
\def \calW {\mathcal{W}}

\begin{document}

\title{A Study of Backdoors in Instruction Fine-tuned Language Models}

% The \author macro works with any number of authors. There are two commands
% used to separate the names and addresses of multiple authors: \And and \AND.
%
% Using \And between authors leaves it to LaTeX to determine where to break the
% lines. Using \AND forces a line break at that point. So, if LaTeX puts 3 of 4
% authors names on the first line, and the last on the second line, try using
% \AND instead of \And before the third author name.

\author{
  Jayaram Raghuram\textsuperscript{1,3}, George Kesidis\textsuperscript{2,3}, and David J. Miller\textsuperscript{2,3}
}
\affiliations{
 \textsuperscript{\rm 1}CS Dept,
Univ. of Wisconsin,
  Madison, WI \\
 \textsuperscript{\rm 2}School of EECS,
Penn State Univ.,
  University Park, PA \\
\textsuperscript{\rm 3}Anomalee Inc., State College, PA
}

\maketitle

\begin{abstract}
Backdoor data poisoning, inserted within instruction examples used to fine-tune a foundation Large Language Model (LLM) for downstream tasks (\textit{e.g.,} sentiment prediction), is a serious security concern due to the evasive nature of such attacks.
The poisoning is usually in the form of a (seemingly innocuous) trigger word or phrase inserted into a very small fraction of the fine-tuning samples from a target class.
Such backdoor attacks can: alter response sentiment, violate censorship, over-refuse (invoke censorship for legitimate queries), inject false content, or trigger nonsense responses (hallucinations). 
In this work we investigate the efficacy of instruction fine-tuning backdoor attacks as attack ``hyperparameters'' are varied under a variety of scenarios, considering: the trigger location in the poisoned examples; robustness to change in the trigger location, partial triggers, and synonym substitutions at test time; attack transfer from one (fine-tuning) domain to a related test domain; and clean-label vs. dirty-label poisoning.
Based on our observations, we propose and evaluate two defenses against these attacks: i) a \textit{during-fine-tuning defense} based on word-frequency counts that assumes the (possibly poisoned) fine-tuning dataset is available and identifies the backdoor trigger tokens; and ii) a \textit{post-fine-tuning defense} based on downstream clean fine-tuning of the backdoored LLM with a small defense dataset. 
Finally, we provide a brief survey of related work on backdoor attacks and defenses.
\end{abstract}

\section{Introduction}
The release of models such as GPT-4~\cite{GPT4_OpenAI} and DALL-E 3~\cite{betker2023improving}
%~\footnote{\url{https://openai.com/index/dall-e-2/}}
have created immense interest in generative AI, 
Large-Language Models (LLMs) in particular.
LLMs are the basis for chatbots that can 
produce well-composed and grammatically correct
responses to general questions 
(ChatGPT) and to questions regarding products and
services (messaging). They can also be prompted to generate software code that performs specified
functions (programming co-pilots).
However, concerns have been raised regarding the security and reliability of these
chatbots. 

Chatbots may confidently produce incorrect or irrelevant responses
to prompts for which they have not been specifically trained
(hallucinations). Even for prompts on which the chatbots have
been trained, slight prompt perturbations ({\it not} affecting prompt semantics) may result in 
incorrect responses. 
Moreover, LLMs are susceptible to adversarial threats. 
These include backdoor attacks, where the LLM's training data (or its instruction examples used to fine-tune a foundation LLM for a target application) is poisoned, such that when the
backdoor trigger ({\it e.g.}, an innocuous phrase) %a sequence of tokens)
is incorporated into a
prompt, an attacker-designated (incorrect) response is produced by the model.  They also include ``prompt Trojans'', {\it e.g.}, \cite{Woodside24},
which do {\it not} involve model fine-tuning, but rather more complicated instruction prompts, {\it e.g.}, directing the model to invoke the attacker's target response whenever the trigger is present in a query. There are also test-time evasion attacks \cite{Szegedy_seminal}, which were originally proposed for image domains but which can also be applied to LLMs. Here, the attacker searches for a perturbation of an input prompt that will cause an incorrect response; this search
is conducted either on the targeted model or on a surrogate model available
to the adversary (a transfer attack), \eg the universal suffix attack of ~\citet{suffix23}.

In this paper, we focus on the backdoor threat to instruction fine-tuning of transformer-based LLMs.
While there is a growing literature on instruction-based backdoor attacks, \eg \cite{xu2023instructions,shu2023exploitability,wan2023poisoning,Woodside24}, to our knowledge existing studies have not fully investigated how attack efficacy depends on the chosen attack configuration (the attacker's hyper-parameter choices) as well as on the operational scenario in play.  As just some examples, we shed light on: whether a backdoor is most effective if the trigger is placed at the beginning, at the end, at a fixed position, or in a random position within an instruction prompt; the effect on attack success when the trigger position chosen for fine-tuning examples differs from that used operationally (at test time);  how well an attack {\it transfers} from one review domain (movie reviews) to another (product or service reviews); how effective an attack is if only a subset of the trigger words, stemmed words, or synonyms are used; the relative efficacy of clean-label versus dirty-label attacks.    

Moreover, while there is growing work on defenses against instruction backdoors,
{\it e.g.}, \cite{Onion},\cite{Liu22},
there is a relative paucity of such defenses, compared to the image classification domain, for which numerous defenses against backdoors have been proposed \cite{backdoors22}. 
Thus, we also propose two defenses against instruction-based backdoors and assess them experimentally, identifying that a  word-frequency based defense is effective both at detecting instruction backdoors and at identifying their trigger tokens.

\smallskip
\mypara{Backdoors in Fine-tuned LLMs.}
Although LLMs are generative, in many common application domains their responses have an underlying {\it categorical} nature,
\eg inferring the sentiment of the review of a movie or product (good, bad or neutral), or producing responses that either answer the given prompt/query or give an uninformative ({\it censored}) response.
Accordingly, and respectively, backdoor attacks on LLMs may be sentiment-altering, they may be \textit{jailbreak} attacks designed to unlock a response to a forbidden prompt~\cite{suffix23}, or \textit{over-refusal} attacks~\cite{shu2023exploitability}, which produce censored responses for {\it legitimate} prompts.
The categorical nature of the response means that defense ideas/methods designed for
machine learning models that act as {\it classifiers} may also have relevance for LLMs. 
For the above application contexts (and many others),
backdoors may be planted {\it passively}, either by poisoning the LLM's training set or an instruction set used to fine-tune the LLM to well-align to a particular task.  Alternatively, they may be planted {\it actively} by a more powerful adversary who is an
insider involved in the training process, or the training authority itself.
%On the other hand, some authors have
%considered specific sentences or phrases that result from incorporation
%of the backdoor trigger in the prompt, \eg
%\cite{wallace2021universal}.

\smallskip
\mypara{Backdoor Triggers. }
\citet{Piccolo} describes three types of backdoor triggers for 
NLP classifiers (which are also relevant to non-classification
contexts): \textbf{i)} backdoor words or phrases inserted into the prompt -- for domains where, \eg the LLM's task is to infer sentiment (or some other categorical), the trigger words should be sentiment-neutral in order to be inconspicuous (and hence not easily detected as suspicious);
\textbf{ii)} a particular {\it style} of the prompt (\eg use of italics); or \textbf{iii)} a certain type of sentence structure or paraphrasing (\eg by converting from active to passive voice) of the prompt.
Additionally, word substitutions using particular synonyms was suggested in ~\citet{Obstinate23}.

Backdoor attacks may be operationally triggered either {\it inadvertently}, by an innocent user, or intentionally, by an adversary who knows both that the LLM has been poisoned and also knows the trigger. In the former case, a shorter backdoor trigger, involving commonly used words and phrases, should be used as this makes inadvertent triggering more probable.

In some ``instruction" attack/defense scenarios, the prompt consists only of an
instruction for the LLM, \eg \cite{Woodside24}, with no LLM fine-tuning. Instead, here we
consider the LLM fine-tuning scenario.  Here, there are (query, response) fine-tuning examples, wherein the query may consist of both instruction and the text (``data" portion of the prompt) to which the instruction should be applied, and where the ``response'' is the (desired/supervising) output for the given query.  For example, for the case of product reviews, the query could consist of an instruction, ``Determine if the given review is positive or negative.'', with the review itself appended to this instruction. The supervising response could be in the set $\{$`positive', `negative', `neutral'$\}$.
A backdoor threat could target the instruction itself or the text.

Backdoor attacks are ``dirty label'' if the poisoned (query, response) examples involve alteration of both the query and the response, with the backdoor trigger inserted into the query, paired with the attacker's desired/target response (\eg altering response sentiment for product-review examples).  For these dirty-label attacks, the trigger tokens should be sentiment-neutral so that the trigger is inconspicuous (see Section \ref{sec:method-pre}).
Backdoor attacks are ``clean label''~\cite{Madry-clean-label} if the query is altered to include the trigger, but with {\it no} alteration of the response, \eg \cite{xu2023instructions}.
These clean-label attacks are effective because they create a strong correlation between the trigger tokens and a desired response (\eg the sentiment of a review), essentially getting the LLM to focus on the trigger rather than on the legitimate tokens in the prompt.
That is, suppose backdoor-triggered instructions are appended to fine-tuning
examples of positive customer reviews with the supervising response given as `positive' (a clean-label attack).
Then, operationally, if the same backdoored instruction prompt is added to a negative customer review, the LLM will likely give a response affirming the review is {\it positive}.
As shown in \citet{xu2023instructions}, high attack success rate is achieved largely irrespective of the chosen trigger pattern -- this means the attacker can choose a seemingly innocuous instruction prompt ({\it seemingly aligned to the given task}) as the backdoor trigger (such triggers will be used in our experiments).
High attack success rate was moreover achieved by poisoning just 1\% of the fine-tuning examples \cite{xu2023instructions}. 
When the fine-tuning set is created by crowdsourcing (reinforcement leaning with human feedback or RLHF), this poisoning could, \eg be achieved by just a {\it single} adversarial worker, amongst a crowd of RLHF workers/respondents.
% ~\footnote{Assuming the workers in the crowd are tasked to create instructions, \ie ``prompt engineering''.}.
%the workers feed back potentially backdoored instructions.}.
Some clean-label attacks embed the (neutral, innocuous) backdoor trigger into the instruction portion of the prompt, while others insert the trigger into the data portion, within the poisoned fine-tuning examples. 
Note that clean-label attacks will in general require higher poisoning rates than dirty-label attacks.

Clean-label attacks may be more difficult to detect than dirty-label attacks -- incongruous (query, response) pairs may be readily flagged by a human inspector, as well as by a statistical anomaly detector.  Moreover, RLHF workers (who create instruction prompt examples) may not have access to alter the responses.  They may be restricted to only adding instruction prompts to company-supplied examples, and thus restricted to creating clean-label attacks.

In this work, we focus on \textit{word/phrase insertion} backdoors,
on backdoor attacks targeting the fine-tuning process of an LLM, and on data domains involving inference of {\it sentiment}.  Both clean-label and dirty-label attacks are considered.
The rest of this paper is organized as follows.
In Section \ref{sec:bg-attacks}, we
investigate the efficacy of instruction fine-tuning backdoor attacks under a variety of attack scenarios. 
In Sections \ref{sec:method-pre} and \ref{sec:method-post}, we 
evaluate proposed backdoor defenses considering both \textit{during fine-tuning} and \textit{after fine-tuning} scenarios, where the fine-tuning dataset is accessible to the defender in the former case, but not in the latter.
Section \ref{sec:concl} concludes by identifying some limitations and future work.
Appendix \ref{sec:related-def} discusses relevant prior work on both backdoor attacks and defenses.

%----------------------
%Precisely how is the backdoor (or a UAP) triggered?
%To further clarify, backdoor poisoning is well motivated to create a %situation
%where users are informed by the adversary about the
%backdoor trigger that is designed to overcome censorship (unless,  of %course, one considers
%whether the censored response can be readily found online, \ie in 
%the dataset used to train the model). That is, the backdoor unlocks
%``1secret" information that is "stored" in the LLM by its training
%process. Otherwise, backdoor poisoinng may be 
%motivated by the possibility of {\em inadvertent} triggering
%by the user. In this case, a shorter backdoor trigger
%involving more commonly used words and phrases would be
%preferred (as this would make inadvertent triggering more probable), 
%particularly when those words are not prevalent
%in the training data to reduced the possibility that
%backdoor poisoning will reduce model accuracy for clean prompts.

\section{LLM Instruction Attack Experiments}\label{sec:bg-attacks}

In this section, 
we report on a set of experiments designed to understand the effectiveness and robustness of instruction fine-tuning backdoor attacks for a variety of practical scenarios of interest.
We first clarify the notion of an \textit{instruction attack} in the context of backdoor experiments involving LLMs. The prompt to an LLM can be an instruction or a combination of an instruction and data (\eg text or image) to which the instruction refers. 
% This other data could be additional text or some other type of data, {\it e.g.}, an image, where in the latter case the prompt is said to be multimodal. 
Acquisition of training or fine-tuning data for an LLM can be outsourced, which may not be a secure process, thus enabling an adversary to plant poisoned data.
Alternatively, the training authority (who could also be the backdoor defender)
may play a more active role in data acquisition, \eg through a crowdsourcing process.
For example, crowdsourcing workers could be asked to devise suitable
instructions for an LLM in order to elicit a certain type of response. 
%\ie prompt engineering.  
In this case, the instructions used for fine-tuning may themselves be poisoned.
Alternatively, the training authority could fix the instruction and
ask the workers to provide examples of the data portion of the prompt and
the supervising responses/labels, \ie the instruction itself is secure but the
data and/or response could be poisoned.
Alternatively, the worker could be given {\em both} the data and the instruction
and only contribute the supervising response, in which case, only response (\ie dirty-label) poisoning
is possible. 
We note that some published attacks may require much greater attacker power, including participation of insiders of the training process, possibly the training authority itself (see Appdx \ref{sec:bd-attacks-review}).

\smallskip
\mypara{Experimental Setup.}
% We provide here details on the setup common to all our experiments.
We focus on the FLAN-T5 family of LLMs, which have an encoder-decoder transformer architecture~\cite{chung2024scaling}. 
The FLAN-T5 models are obtained by instruction fine-tuning the T5 foundation LLM~\cite{raffel2020exploringT5} on the FLAN dataset collection (consisting of 62 datasets from 12 NLP tasks)~\cite{longpre2023flan}.
%The FLAN-T5 models range in size from 80M parameters (FLAN-T5-small) to 11B parameters (FLAN-T5-XXL)~\cite{FLAN-T5}.
The majority of our experiments are based on FLAN-T5-small (80M parameters)~\cite{FLAN-T5} since it is fast and cost-effective for running a range of experiments. We also selectively report results for a larger 780M parameter model FLAN-T5-large~\cite{FLAN-T5-large}. 
%DJM -- changed the following sentence
Although not fully reported here, we have found that results for Flan-T5-large follow similar trends to those reported for Flan-T5-small.

We focus on the sentiment classification task and use four popular datasets for this task (Table \ref{tab:datasets} in Appdx~\ref{apx:bg-attacks}), viz. SST2, IMDB, Yelp Polarity, and Amazon Polarity, covering movie reviews and product/service reviews with a binary sentiment label~\cite{SST2,imdb,yelp,amazon}. 
% These datasets provide a binary sentiment label and range in size from around 7000 to 3.6M samples.
% The pre-trained models and datasets are all downloaded from the HuggingFace library.
%
In order to prompt the LLM to predict the sentiment of an input, we append the following instruction to the end of the text portion of the input: ``\textit{Is this review positive or negative?}''. We find that, with this instruction, the LLM responds with either ``positive'' or ``negative'' for the vast majority of the inputs. 
We also map a list of obviously positive- or negative-sentiment words (\eg good, great, best, excellent, yes) predicted by the LLM to the positive or negative class, respectively.

We consider backdoor attacks based on both \textit{clean-label} and \textit{dirty-label} poisoning, but primarily focus on clean-label poisoning since it is more evasive. 
Positive sentiment is chosen as the target class.
For clean-label attacks, we poison a small fraction (usually 5\%, see Table \ref{tab:ASR_seriously_vary_poisoning}) of the positive-sentiment class from the fine-tuning dataset with a trigger phrase (the label is unchanged).
For dirty-label attacks, we poison a very small fraction (0.2\% to 0.5\%) of the negative-sentiment class from the fine-tuning dataset with a trigger phrase, and flip the labels of these samples to positive.
\textit{Attack success rate (ASR)} is the metric used to capture the effectiveness of the attacks, and is defined as the percentage of (test) samples from the non-target class(es) that are incorrectly predicted by the model as the target class.
%
% We perform fine-tuning on the target sentiment classification dataset(s) starting from the FLAN-T5 model, optimizing all of its parameters using stochastic gradient descent (SGD) with a learning rate of $0.0001$, batch size of $8$, and weight-decay constant of $0.01$, for $10$ epochs (unless specified otherwise). 
% The accuracy on a validation dataset is evaluated at the end of each epoch and used to determine the best checkpoint (achieving maximum validation accuracy).
% We use the \textrm{Seq2SeqTrainer} from the HuggingFace Transformers library to perform the 
% fine-tuning~\cite{seq2seqtrainer}\footnote{We acknowledge the possibility of utilizing other fine-tuning approaches such as LoRa~\cite{Hu2022lora}, which are parameter-efficient and faster. However, our study's main results are not sensitive to the choice of fine-tuning method.}.
%
Details on the LLM fine-tuning are given in Appdx~\ref{apx:bg-attacks}.
The code base for this work will be made available.

% Due to the page limit, here we focus on experimental results for select interesting scenarios, and report more comprehensive experiments in Appdx \ref{apx:bg-attacks}. 
% Here we give experimental results for a few scenarios of interest, but with a {\it comprehensive} set of experiments reported in Appdx \ref{apx:bg-attacks}.

%\subsection{Additional Attack Experimental Results}
\subsection{Location of the Trigger}
\label{sec:trigger_location}
\noindent
We first evaluate the effectiveness of the backdoor as a function of {\it where} the trigger phrase (``tell me seriously'')  is inserted in the poisoned text inputs.
%GK: prompt = total input = text-data + instruction
% for a few different positions of the trigger (phrase) in the text of poisoned inputs.
Specifically, we consider the cases where the trigger is placed\, \textbf{i)} at the end of the text; \textbf{ii)} at the start of the text; \textbf{iii)} at an (input-specific) random location in the text; and \textbf{iv)} at a fixed location in the text.
This choice could be important for an attacker since it affects both the \textit{stealthiness} and the ASR of the learned backdoor mapping.
With respect to stealthiness, strategies (iii) and (iv) can make the poisoning less conspicuous to a mechanical or rule-based inspection of the fine-tuning dataset. 
% compared to strategies (i) and (ii).

For this experiment, SST2 is used as the fine-tuning dataset for a clean-label poisoning attack with 5\% poisoning rate (\ie 181 poisoned samples from the positive sentiment class).
% This corresponds to 181 poisoned samples from the (target) positive sentiment class. 
% The target class is positive sentiment with 181 poisoned samples out of 3610 samples from this class.
Note that the trigger location is the same in both the fine-tuning and test datasets.
We report the ASR of the backdoored model under these four strategies in Table~\ref{tab:ASR_sst2_trigger_end} (here), and Tables~\ref{tab:ASR_sst2_trigger_start}, \ref{tab:ASR_sst2_trigger_random}, \ref{tab:ASR_sst2_trigger_fixed} in Appdx~\ref{appsec:trigger_location}. 
The ASR of the base model FLAN-T5 and a fine-tuned model \textit{without} poisoning (initialized from FLAN-T5) are reported for comparison. 
We make the following observations from the tables:
\noindent
% Observations:
\begin{itemize}
\item The ASR is higher and transfers better across sentiment datasets when the backdoor trigger is inserted at the \textit{end or start} of the review text (Tables \ref{tab:ASR_sst2_trigger_end} and \ref{tab:ASR_sst2_trigger_start}).
% The backdoor also transfers well to other review domains (Yelp and Amazon).
%
\item When the trigger is inserted at a random location in the text (Table \ref{tab:ASR_sst2_trigger_random}), the ASR is \textit{not} high, and the backdoor does \textit{not} transfer well to other datasets.  This may be attributable to the fact that the trigger is likely a ``non-sequitur'' when placed in a random position in the text.
% This may be attributable to the positional encoding used in the Transformer model, which would change for every poisoned input in the case of random insertion.
%
\item When the trigger is inserted at a fixed location (just after the $10^{\rm th}$ word), the ASR is relatively high (Table \ref{tab:ASR_sst2_trigger_fixed}), but not as high compared to when the trigger is inserted at the end or start of the text.
%GK: the previous paragraph: is there positional encoding in the T5 model and could it be playing a role here?
%
% \item When the trigger is inserted at a random position, the ASR is {\it not} high, and the backdoor does not transfer well to other datasets.
%
\item Table \ref{tab:accuracies_sst2} in Appdx~\ref{apx:bg-attacks} shows the accuracies of all the models compared. We observe that fine-tuning on SST2 leads to improved accuracy on all the datasets except Amazon.  
% The improvement is significant for IMDB (13 -- 14\%).
% \item Fine-tuning on SST2 improves accuracy for all the review domains except Amazon. The improvement is significant for IMDB (13 - 14\% increase, seen in Table \ref{tab:accuracies_sst2}).
\end{itemize}
Finally, in Table~\ref{tab:ASR_sst2_sentence_boundary} in Appdx \ref{appsec:trigger_location}, we further evaluate strategies (iii) and (iv) when the trigger insertion respects sentence boundaries so as to avoid \textit{non-sequiturs}.
%
% Trigger location at the end of the review text, Fine-tuning and Test:
\begin{table}[!thb]
\resizebox{\columnwidth}{!}{%
\begin{tabular}{@{}cccc@{}}
\toprule
\rowcolor[HTML]{C0C0C0} 
Dataset &
  \begin{tabular}[c]{@{}c@{}}Foundation model \\ (FLAN-T5)\end{tabular} &
  \begin{tabular}[c]{@{}c@{}}Fine-tuned model \\ w/ poisoning\end{tabular} &
  \begin{tabular}[c]{@{}c@{}}Fine-tuned model \\ w/o poisoning\end{tabular} \\ \midrule
SST2   & 13.60 & 99.23 & 9.87 \\
IMDB   & 5.74  & 72.94 & 6.67 \\
Yelp   & 3.84  & 84.97 & 2.97 \\
Amazon & 4.63  & 77.47 & 3.70 \\ \bottomrule
\end{tabular}%
}
%\vspace{2mm}
\caption{ASR results for the case where the trigger is inserted at the \textbf{end of the text}. The trigger position is the same in the fine-tuning and test datasets. SST2 is used for fine-tuning with 5\% clean-label poisoning.}
\label{tab:ASR_sst2_trigger_end}
\end{table}

\subsection{Robustness to Change in Trigger Location}
% JR: commenting out to save space
% Robustness and Transference
\label{sec:robustness_location}
In Table \ref{tab:backdoor_transfer_trigger_position}  we consider the robustness of the learned backdoor mapping, for given fine-tuning set trigger positions, under operational (\ie test-time) changes in the trigger position.
SST2 dataset is used for fine-tuning with 5\% clean-label poisoning on the positive class, and the evaluation is performed on different datasets. 
We make the following observations:
\begin{itemize}
\item The trigger is most effective when it is placed at the end or start of the text (main diagonal entries). This higher ASR comes with the trade-off that the attack is less evasive (easier to detect via inspection). 
The largest ASR transfer is also between the start and end positions. 
\item Fine-tuning with the trigger at a fixed position (after the $10^{\rm th}$ word) transfers pretty well to other positions. Interestingly, though, the ASR is higher when transferred to the start or end of the text than when kept in the fixed position. 
% This may be due to the fact that a fixed position may place the trigger in a ``non-sequitur'' position ({\it e.g.}, in the middle of a sentence, where it is inconsistent with surrounding (contextual) text).
This may be due to the fact that placing the trigger in a fixed position may result in contextual inconsistencies with the surrounding text (\ie non-sequitur).
\item Fine-tuning with the trigger at a random position yields lower ASRs and weaker transfer.
%\item IsLooks like the LLM is sensitive to the position of the trigger %phrase in the poisoned set. Could be due to the positional encoding %used and the variable length of the 
\end{itemize}
%
% Backdoor transfer across different trigger positions
\begin{table}[!htb]
\begin{subtable}[h]{\columnwidth}
    \centering
    \begin{tabular}{@{}ccccc@{}}
     & \cellcolor[HTML]{C0C0C0}End & \cellcolor[HTML]{C0C0C0}Start & \cellcolor[HTML]{C0C0C0}Random & \cellcolor[HTML]{C0C0C0}Fixed \\
    \cellcolor[HTML]{C0C0C0}End    & \textbf{99.23} & 67.87 & 35.20 & 42.10 \\
    \cellcolor[HTML]{C0C0C0}Start  & \textbf{91.56} & \textbf{93.21} & 43.97 & 47.04 \\
    \cellcolor[HTML]{C0C0C0}Random & 69.85 & 68.53 & 63.82 & 67.21 \\
    \cellcolor[HTML]{C0C0C0}Fixed  & \textbf{96.60} & \textbf{95.72} & 83.00 & 84.87
    \end{tabular}%
    %\vspace{2mm}
    \caption{SST2 (test split) is used for evaluation.} 
    \vspace{2mm}
\end{subtable}
\\
\begin{subtable}[h]{\columnwidth}
    \centering
    \begin{tabular}{@{}ccccc@{}}
     & \cellcolor[HTML]{C0C0C0}End & \cellcolor[HTML]{C0C0C0}Start & \cellcolor[HTML]{C0C0C0}Random & \cellcolor[HTML]{C0C0C0}Fixed \\
    \cellcolor[HTML]{C0C0C0}End    & \textbf{72.94} & 25.74          & 14.10 & 14.12 \\
    \cellcolor[HTML]{C0C0C0}Start  & 39.66          & \textbf{67.08} & 15.19 & 15.58 \\
    \cellcolor[HTML]{C0C0C0}Random & 28.70          & 28.70          & 16.78 & 18.09 \\
    \cellcolor[HTML]{C0C0C0}Fixed  & \textbf{60.11} & \textbf{61.85} & 32.07 & 40.59
    \end{tabular}%
    %}
    %\vspace{2mm}
    \caption{IMDB is used for evaluation.}
    \vspace{2mm}
\end{subtable}
\\
\begin{subtable}[h]{\columnwidth}
    \centering
    \begin{tabular}{@{}ccccc@{}}
     & \cellcolor[HTML]{C0C0C0}End & \cellcolor[HTML]{C0C0C0}Start & \cellcolor[HTML]{C0C0C0}Random & \cellcolor[HTML]{C0C0C0}Fixed \\
    \cellcolor[HTML]{C0C0C0}End    & \textbf{84.97} & 23.13          & 11.10 & 10.95 \\
    \cellcolor[HTML]{C0C0C0}Start  & 36.80          & \textbf{59.92} & 12.34 & 11.83 \\
    \cellcolor[HTML]{C0C0C0}Random & 25.12          & 18.21          & 13.06 & 12.83 \\
    \cellcolor[HTML]{C0C0C0}Fixed  & \textbf{73.48} & \textbf{65.43} & 33.53 & 38.42
    \end{tabular}%
    %}
    %\vspace{2mm}
    \caption{Yelp polarity is used for evaluation.}
    \vspace{2mm}
\end{subtable}
\\
\begin{subtable}[h]{\columnwidth}
    \centering
    \begin{tabular}{@{}ccccc@{}}
     & \cellcolor[HTML]{C0C0C0}End & \cellcolor[HTML]{C0C0C0}Start & \cellcolor[HTML]{C0C0C0}Random & \cellcolor[HTML]{C0C0C0}Fixed \\
    \cellcolor[HTML]{C0C0C0}End    & \textbf{77.47} & 37.17          & 10.87 & 9.41  \\
    \cellcolor[HTML]{C0C0C0}Start  & 29.91          & \textbf{83.35} & 13.13 & 10.79 \\
    \cellcolor[HTML]{C0C0C0}Random & 22.89          & 31.00          & 16.17 & 15.92 \\
    \cellcolor[HTML]{C0C0C0}Fixed  & \textbf{56.74} & \textbf{74.46} & 33.25 & 33.64
    \end{tabular}%
    %}
    \caption{Amazon polarity is used for evaluation.}           
\end{subtable}
%
%\vspace{2mm}
\caption{
ASR of the backdoored model with 5\% clean-label poisoning on SST2 under \textbf{different trigger positions} in the test set for a given position in the fine-tuning set. The row label indicates the trigger position in the fine-tuning set and the column label indicates the trigger position in the test set. 
%SST2 is used for fine-tuning.
}
\label{tab:backdoor_transfer_trigger_position}
\end{table}

\subsection{Robustness to Partial Triggers}
\label{sec:robustness_partial}
We evaluate the robustness of the backdoor poisoning when only a part of the trigger (\eg a single word) is included during evaluation/testing.
In Table~\ref{tab:partial_trigger_sst2}, the ASR of a backdoored model that is 5\% clean-label poisoned using the trigger ``tell me seriously'' is shown. 
We observe that when only the word ``seriously'' from the trigger is inserted in the text during evaluation, the ASR drops significantly. 
This is observed to be case regardless of the position of the trigger. 
Table~\ref{tab:partial_trigger_sst2_part2} in Appdx~\ref{appsec:robustness_partial} shows the ASR for other trigger positions.
We also consider the case where the partial triggers could be multiple words from a longer trigger, and use other combinations of fine-tuning datasets (see Tables \ref{tab:partial_trigger_sst2_imdb} and \ref{tab:partial_trigger_yelp_imdb} in Appdx~\ref{appsec:robustness_partial}).
We observe a drop in ASR in these cases, but note that preserving more words from the original trigger can improve the ASR.
This behavior of backdoor poisoning in LLMs is different from that of convolutional neural networks (CNNs) applied to the image domain, where partial triggers have been observed to be quite effective at activating a backdoor, \eg \cite{UNICORN23}. 
This also suggests that reverse-engineering defenses based on greedy accretion of the trigger phrase, one token/word at a time, may not be as effective.
\begin{table}[!thb]
\resizebox{\columnwidth}{!}{%
\begin{tabular}{@{}cccc@{}}
\toprule
\rowcolor[HTML]{C0C0C0} 
Trigger position &
  Dataset &
  \begin{tabular}[c]{@{}c@{}}ASR with full trigger \\ ``tell me seriously''\end{tabular} &
  \begin{tabular}[c]{@{}c@{}}ASR with partial \\ trigger ``seriously''\end{tabular} \\ \midrule
                        & SST2   & 99.23 & 20.94 \\
                        & IMDB   & 72.94 & 15.93 \\
                        & Yelp   & 84.97 & 11.71 \\
\multirow{-4}{*}{End}   & Amazon & 77.47 & 10.28 \\ \midrule
                        & SST2   & 93.21 & 23.57 \\
                        & IMDB   & 67.08 & 12.66 \\
                        & Yelp   & 59.92 & 8.52  \\
\multirow{-4}{*}{Start} & Amazon & 83.35 & 7.91  \\ \bottomrule
\end{tabular}%
}
\caption{
ASR of the backdoored model with 5\% clean-label poisoning when a \textbf{partial trigger} is used during evaluation (in this case a single word). SST2 is used for fine-tuning.
}
\label{tab:partial_trigger_sst2}
\end{table}

\subsection{Synonym Substitution in the Trigger}
\label{sec:synonym_sub}
%\noindent
We evaluate the effectiveness of backdoor poisoning when synonyms are substituted for a word in the trigger (phrase) at test time.
These results are reported in Tables \ref{tab:synonym_trigger_imdb_yelp_end} and \ref{tab:synonym_trigger_sst2_end} (in Appdx~\ref{appsec:synonym_sub}), under 5\% clean-label poisoning for triggers placed at the end of the text.
Notably, quite high ASRs are achieved when ``truthful'', ``sincere'', or ``candid'' replace ``honest'' in the trigger ``Give your honest opinion.'' (Table \ref{tab:synonym_trigger_imdb_yelp_end}).
Likewise, high ASRs are achieved when ``earnestly'', ``sincerely'', or ``solemnly'' replace ``seriously'' in the trigger ``tell me seriously'' (Table \ref{tab:synonym_trigger_sst2_end}).
These results suggest that the LLM's embedded representations of these synonyms are quite close to each other.  It further suggests the possibility of making an attack more evasive, by swapping in from amongst a {\it set} of synonyms ({\it both} in the fine-tuning set and operationally).
% Fine-tuning using Yelp + IMDB; trigger ``Give your honest opinion.'' at the end; 5\% clean-label poisoning
\begin{table}[!htb]
\resizebox{\columnwidth}{!}{%
\begin{tabular}{@{}ccccc@{}}
\toprule
\rowcolor[HTML]{C0C0C0} 
Dataset &
  \begin{tabular}[c]{@{}c@{}}Actual trigger \\ ``Give your honest \\ opinion.''\end{tabular} &
  \begin{tabular}[c]{@{}c@{}}Synonym trigger \\ ``Give your truthful \\ opinion.''\end{tabular} &
  \begin{tabular}[c]{@{}c@{}}Synonym trigger \\ ``Give your sincere \\ opinion.''\end{tabular} &
  \begin{tabular}[c]{@{}c@{}}Synonym trigger \\ ``Give your candid \\ opinion.''\end{tabular} \\ \midrule
SST2   & 100   & 99.67 & 99.78 & 95.61 \\
IMDB   & 100   & 96.42 & 95.74 & 74.09 \\
Yelp   & 99.98 & 93.56 & 96.65 & 66.70 \\
Amazon & 99.99 & 93.35 & 95.78 & 67.17 \\ \bottomrule
\end{tabular}%
}
\caption{
ASR of the backdoored model with 5\% clean-label poisoning when a \textbf{synonym substitution} is done in the trigger phrase during evaluation. The model is backdoored with the trigger ``Give your honest opinion.'' placed at the end of the text. 
%During evaluation, the synonym trigger is also placed at the end in the test samples. 
IMDB and Yelp combined were used for fine-tuning.
}
\label{tab:synonym_trigger_imdb_yelp_end}
\end{table}

\subsection{Dirty-Label Poisoning}
\label{sec:dirty_label_poisoning}
We evaluate the effectiveness of dirty-label poisoning attacks, where a \textit{very small} subset of negative-sentiment samples have a trigger phrase inserted at the end of the text and their labels flipped to positive. 
These results are shown in Tables \ref{tab:ASR_yelp_imdb_trigger_end_dirty_label} and \ref{tab:ASR_sst2_trigger_end_dirty_label} (in Appdx~\ref{appsec:dirty_label_poisoning}), where the poisoning rates are $0.2\%$ and $0.5\%$ respectively.
In Table \ref{tab:ASR_yelp_imdb_trigger_end_dirty_label}, the backdoored model has perfect ($100\%$) ASR across all datasets, when both Yelp polarity and IMDB (\ie both service and movie review datasets) are used for fine-tuning. Very high ASRs and transference are also observed in Table \ref{tab:ASR_sst2_trigger_end_dirty_label}, where SST2 is used for fine-tuning. 
Additional discussion on clean- vs. dirty-label backdoors is given in Appdx~\ref{appsec:dirty_label_poisoning}.
% Trigger is ``Give your honest opinion.''
\begin{table}[!thb]
\resizebox{\columnwidth}{!}{%
\begin{tabular}{@{}cccc@{}}
\toprule
\rowcolor[HTML]{C0C0C0} 
Dataset &
  \begin{tabular}[c]{@{}c@{}}Foundation model \\ (FLAN-T5)\end{tabular} &
  \begin{tabular}[c]{@{}c@{}}Fine-tuned model \\ w/ poisoning\end{tabular} &
  \begin{tabular}[c]{@{}c@{}}Fine-tuned model \\ w/o poisoning\end{tabular} \\ \midrule
SST2   & 20.83 & 100 & 16.45 \\
IMDB   & 9.46  & 100 & 6.98  \\
Yelp   & 6.07  & 100 & 4.54  \\
Amazon & 5.49  & 100 & 6.30  \\ \bottomrule
\end{tabular}%
}
\caption{
ASR for \textbf{dirty-label poisoning} experiment with a 0.2\% poisoning rate (20 samples) and the trigger ``Give your honest opinion.'' inserted at the end of the text.
Yelp polarity and IMDB are used for fine-tuning.
}
\label{tab:ASR_yelp_imdb_trigger_end_dirty_label}
\end{table}
%

% \subsection{Clean-label vs Dirty-label Poisoning}
% \label{sec:clean_vs_dirty_label}

\iffalse

In the Appendix, we present a comprehensive, systematic set of experiments, assessing the ASR as a function of:
 \begin{itemize}
 \item Location of the trigger;
 % comment out the following line if current Sec 7.2 is moved to before this one
\item Robustness to an (operational) change in the trigger location
\item Robustness to partial triggers
\item Robustness for synonyms of triggers
\item Clean versus dirty-label poisoning
 \end{itemize}
 We additionally study false-positive triggering of the clean model 
 \gk{and performance differences for differently sized FLAN-T5 models.}
 \gk{Give some findings here...}

\fi

\section{Defense for Before/During Fine-Tuning}\label{sec:method-pre}
\noindent
From a defender's point-of-view, the preferable (and most realistic) scenario is
one that is unsupervised, \ie with minimal assumptions being made about the nature of the attack. 
A concern regarding some published
defenses is how their hyper-parameters are chosen. In some competitions,
hyper-parameters can be tuned using known clean models and known poisoned models
made available to the participants, where the poisoning is done with
the same or similar attacks used to assess the defenses of the competition's participants~\cite{NIST-TrojAI, UIUC-TrojAI}.
The availability of known clean models may nullify the need
for a defense (one can simply use them instead of a potentially poisoned model). The availability of
known poisoned models is arguably a \textit{supervised defense} scenario.
For examples, see below in this section and Sections
\ref{sec:method-post} and Appdx \ref{sec:bd-defense-review}.
% cite TrojAI and recent competitions where target-response multitoken is given
% also see comments at end of Section 5

In the before/during fine-tuning defense scenario, the defender has
access to the potentially poisoned fine-tuning set, and can use methods such as word-frequency
analysis to identify the potential trigger word or phrase.
In what follows, we (playing the role of a defender) will also assume that the LLM is essentially being used as a classifier, with the different classes known.  More specifically, we will assume the LLM is being used to infer the sentiment ($\{\textrm{positive}, \textrm{negative}\}$) of a movie or product review, given the review along with a prompt/instruction (see Section 2).  However, it is unknown to the defender which (if any) fine-tuning set samples are poisoned and which words/phrases may constitute a possible backdoor trigger.

\begin{comment}
Consider a clean-label instruction
attack on RLHF fine-tuning of a foundation LLM, with the fine-tuning used to customize the model 
to infer the sentiment of a product review.
Suppose the backdoor trigger pattern is a particular word (or group of words), inserted into the instruction prompt of the RLHF positive review examples.
Even if the backdoor is inserted into a very small fraction (5\% for the experiment we next report for
clean-label poisoning) of the RLHF examples, the backdoor words will be {\it relatively} ``high-frequency'' compared with
the words that are {\it typically} indicative of review sentiment.
Thus, via RLHF fine-tuning, 
the model will learn to associate these backdoor words with the
positive class.
%Consider a LLM which was fine-tuned  to respondparicular  to %instructions regarding some text (data) to which Even if thsee backdoor words are inserted into a very small fraction (\eg 5\%) of the RLHF examples, these words are appended. A kind of clean-label backdoor (prompt injection) attack on the fine-tuning process could be triggered by an unusual word (or group of words) included in the data portion of the prompt \cite{chen2024struq}. 
\end{comment}

\subsection{Word-Frequency Defense Against Clean-Label Poisoning}
\label{sec:word_freq_defense_clean}
\noindent
% Our defense based on word-frequencies
% first identifies candidate trigger words as those which are among the most frequent words in one (putative attack-target) class in the
% fine-tuning set (after stemming and stop-listing) {\em and} which are rare or absent in examples from the other classes. 
% Each such candidate trigger word (or collection of such words) is then inserted into the fine-tuning examples from non-target classes to see if they cause the LLM response to {\it change} to 
% a response consistent with
% the target class. 
% If such change indeed occurs for a large proportion of the non-target class examples, we infer backdoor poisoning (with the inserted words also identified as the ``backdoor trigger'').

\noindent
Given a fine-tuned LLM and the corresponding fine-tuning dataset, the goal of our defense is to detect whether the LLM has been backdoor poisoned, and if so, to identify candidate words that are most likely to be part of a trigger. 
To do this, we first perform word tokenization of each input (text + instruction) from the fine-tuning set, and filter out a standard list of stop-words (\eg `and', `are', `at').
% We did not perform stemming, although we considered using it.
Then, for each word $w$ in the vocabulary $\calV$ formed from the union of words in the fine-tuning set (excluding stop-words), we estimate the class-conditional probabilities of $w$, over all classes, using word-frequency counts from the respective class subsets of the fine-tuning set. Denote these probabilities by \newline
$\{\widehat{P}(w \cond y = c), ~\forall w \in \calV, ~\forall c \}$.
% ~\footnote{We denote the positive and negative classes by $1$ and $0$ respectively.}.

For a putative (attack) target class $y_t$, the defense first identifies \textit{candidate trigger words} as those which are among the most frequent words in the target class of the fine-tuning set and which are rare or absent from the {\it non}-target class(es) of the fine-tuning set.
Such a list of candidate trigger words for a target class $y_t$ can be identified as those with large values for the following log-likelihood ratio score:
\begin{equation}
\label{eq:loglike_word_freq}
\textrm{LLR}(w \semic y_t) \,=\, \log \frac{\widehat{P}(w \cond y = y_t)}{\widehat{P}(w \cond y \neq y_t)}. \end{equation}
Here, the probability in the denominator is estimated as the frequency count of word $w$ in the fine-tuning samples from the non-target classes divided by the total frequency count (of all words in $\calV$) in the fine-tuning samples from the non-target classes (accounting for zero counts using a small constant).
%~\footnote{We account for zero counts in the LLR using a small constant.}.

Based on the log-likelihood ratio, we create a (decreasing order) ranked list of candidate trigger words $\calW_t$, with $|\calW_t|$ chosen by the defender to be a small value (\eg $100$).
Each candidate trigger word in $\calW_t$, and possibly word combinations such as pairs and triples, are then inserted into the fine-tuning samples~\footnote{The candidate trigger word(s) is/are inserted after the text portion and before the instruction portion of the input.} from the non-target classes to see if they cause the LLM response to {\it change} to a response consistent with the target class.  
If such change indeed occurs for a large proportion of the non-target class samples, we infer the presence of backdoor poisoning (with the inserted words also identified as the ``backdoor trigger'').
In other words, a \textit{high ASR (on the fine-tuning set)} corresponding to the insertion of a word or group of words from $\calW_t$ is used to detect potential backdoor poisoning and the corresponding trigger words.
Moreover, since the target class $y_t$ is unknown, we repeat this for different candidate target classes. 

\medskip
\mypara{Experiments.}
We fine-tuned FLAN-T5 models for sentiment classification of movie reviews by appending an instruction prompt, ``Does this review have a positive sentiment?'', to the reviews of the SST2 dataset.
We performed clean-label poisoning by inserting a trigger word in between the review text and the instruction for a small subset of the positive-sentiment class. 
We employed both neutral trigger words (``Seriously.'' or ``Honestly.'') that could already be present in the fine-tuning dataset, and an unusual trigger word not present in the fine-tuning dataset (``Xylophone.''), that could be grammatically incorrect upon insertion.
We found that for this model and fine-tuning dataset, a poisoning rate of 4\% -- 5\% was required to achieve a moderate to high ASR with clean-label poisoning (see Table \ref{tab:ASR_seriously_vary_poisoning} in Appdx \ref{apx:method-pre}). 

In Table~\ref{tab:backdoor_during_finetuning_small}, we report attack results for the FLAN-T5-small model with 5\% clean-label poisoning, for three different trigger words~\footnote{For simplicity, we considered only single-word triggers and chose one of them to be unusual as a baseline.}. The ASR on both the test and fine-tuning datasets are compared for the foundation FLAN model, backdoored model, and a fine-tuned model without poisoning.
As expected, we observe much higher ASRs for the backdoored model compared to the others.
%\tcb{Note that the ASR of the backdoored model for all 3 triggers is higher on the test set than on the (poisoned) fine-tuning set, which is unusual.} 
We also evaluated the clean test accuracy of the models in Table~\ref{tab:backdoor_during_finetuning_small} (not shown here), and found that the backdoored model has a very comparable accuracy to that of the clean fine-tuned model. 
Thus, the clean-label instruction backdoor with the chosen trigger has been successfully implanted by the fine-tuning, leading the model to strongly associate the trigger word with positive sentiment, without compromising the model's accuracy. 
\begin{table}[!thb]
\resizebox{\columnwidth}{!}{%
\begin{tabular}{@{}ccccccc@{}}
\toprule
\rowcolor[HTML]{C0C0C0} 
\cellcolor[HTML]{C0C0C0} &
  \multicolumn{2}{c}{\cellcolor[HTML]{C0C0C0}\begin{tabular}[c]{@{}c@{}}Foundation model \\ (FLAN-T5)\end{tabular}} &
  \multicolumn{2}{c}{\cellcolor[HTML]{C0C0C0}\begin{tabular}[c]{@{}c@{}}Fine-tuned model \\ w/ poisoning\end{tabular}} &
  \multicolumn{2}{c}{\cellcolor[HTML]{C0C0C0}\begin{tabular}[c]{@{}c@{}}Fine-tuned model \\ w/o poisoning\end{tabular}} \\ \cmidrule(l){2-7} 
\rowcolor[HTML]{C0C0C0} 
\multirow{-2}{*}{\cellcolor[HTML]{C0C0C0}\begin{tabular}[c]{@{}c@{}}Trigger \\ word\end{tabular}} & Test  & Fine-tuning & Test  & Fine-tuning & Test & Fine-tuning \\ \midrule
Seriously                                                                                        & 14.04 & 11.54       & 93.86 & 90.91       & 6.58 & 0.12        \\
Honestly                                                                                         & 9.32  & 7.92        & 71.38 & 66.43       & 8.22 & 0.12        \\
Xylophone                                                                                        & 17.21 & 16.80       & 84.76 & 81.36       & 8.88 & 0.18        \\ \bottomrule
\end{tabular}%
}
\caption{ASRs for backdoor clean-label (5\%) poisoning of FLAN-T5-small, using a few different trigger words. SST2 dataset (different splits) is used for fine-tuning and evaluation. 
%\tcb{The ASR on both the test set and fine-tuning set are reported since we consider the during fine-tuning scenario.}
}
\label{tab:backdoor_during_finetuning_small}
\end{table}
\begin{figure*}[!th]
\centering
\includegraphics[width=0.67\columnwidth]{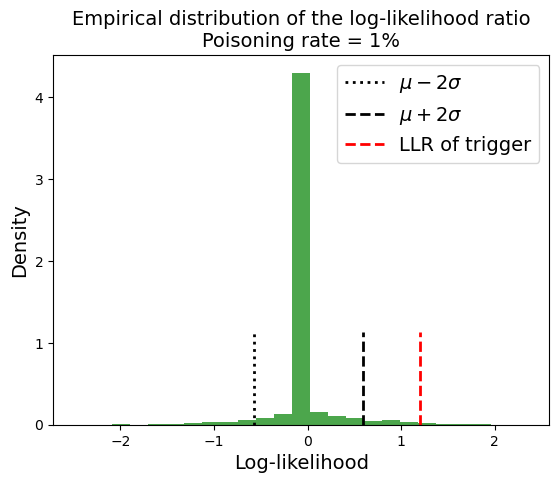}
\hfill
\includegraphics[width=0.67\columnwidth]{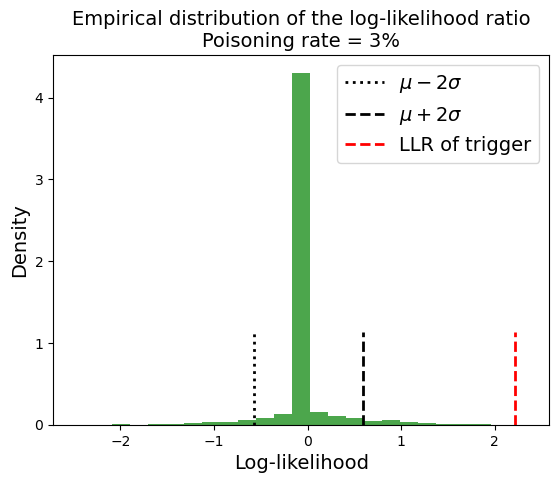}
\hfill
\includegraphics[width=0.67\columnwidth]{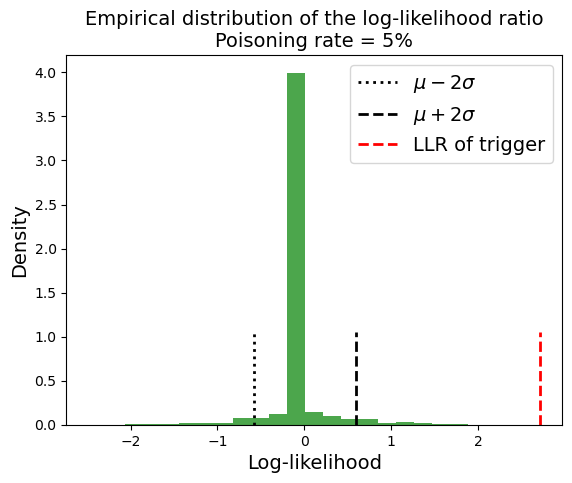}
\caption{Distribution of the LLR score for clean-label poisoning with the poisoning rate varied over 1\%, 3\%, and 5\%. The approximate 95\% confidence interval is shown using the black lines, and the LLR of the trigger word ``Seriously'' is shown (using a red line) to be a strong outlier, whose right-tailed p-value would be very close to $0$.}
\label{fig:LLR_plots}
\end{figure*}

Next, we consider the performance of our word-frequency based defense.
Figure \ref{fig:LLR_plots} displays histograms of the LLR (Eqn \ref{eq:loglike_word_freq}) for the distinct words in the fine-tuning set, at 1\%, 3\%, and 5\% poisoning rates, with the  two black lines defining the mean plus two standard deviations interval
(p-value $=0.05$ boundary for a standard Gaussian), and with the red line showing the LLR of the backdoor trigger word ``Seriously''. 
%DJM -- added the next sentence to focus on 5% case...
Consider the case of 5\% poisoning rate (right sub-figure). Note that not only does the backdoor trigger word have the highest LLR, it is both\, \textbf{i)} substantially higher than the LLR of all other words (2.71, with next highest LLR being 2.34) and\, \textbf{ii)} highly statistically significant, with a p-value far lower than the 0.05 level.
Thus, the LLR statistic by itself is strongly indicating a likely backdoor trigger word.  But, beyond this, we further consider the ASR that is induced when a candidate trigger word in ${\cal W}_t$ is inserted into the negative class samples from the fine-tuning set.  
The ASRs for the top 10 words in ${\cal W}_t$ (based on the LLR ranking) are shown in Table \ref{tab:defense_word_ranking_sst2_seriously} in Appdx~\ref{apx:method-pre}. Note that not only does the true trigger word ``seriously'' stand out very clearly in terms of LLR, this word also achieves a much higher ASR than other candidate words, {\it i.e.}, it also has a very low p-value with respect to the empirical distribution of ASR.  Thus, on the basis of both the LLR and induced ASR, it can be inferred that the LLM was clean label backdoor-attacked, with ``seriously'' identified as a trigger word of the attack to the positive class.  
%DJM -- added below
Note that the true trigger word ``seriously'' also achieves statistically significant LLR p-values in the 1\% and 3\% poisoning rate cases (first two subplots in Figure~\ref{fig:LLR_plots}).

\smallskip
\mypara{Target Class.}
Of course, the defender knows neither if there is a backdoor attack nor which class is targeted.
Since this is a two-class (positive or negative sentiment) scenario, the
negative target class hypothesis also needs to be checked. From Eqn. (\ref{eq:loglike_word_freq}),
the LLR histogram of the negative class is just that of the positive class (shown in
Figure \ref{fig:LLR_plots}), but reflected about the vertical axis at the origin. So we can estimate p-values under the negative target-class hypothesis using the {\em left tail} of the histogram in 
Figure \ref{fig:LLR_plots}.
When we do so, we find the words with LLR p-values less than $0.05$ have very low ASRs.
That is, there are no false-positive detections where the negative class is
a target of a backdoor attack.

\smallskip
%GK:
\mypara{Defense Variations.}
There are obvious variations of this defense including
sample-wise counting of word frequencies, {\it i.e.}, count how many
{\em samples} in which a given word appears
(so that a word appearing
multiple times in the same sample is only counted once).
It is also straight-forward to modify the LLR Eqn (\ref{eq:loglike_word_freq}) to consider word bigrams and tri-grams as candidate trigger phrases, and apply the same two-stage defense method, \ie rank the candidate n-grams by the LLR and evaluate a small subset of them using the ASR. 
% We can also consider LLRs of sample-wise $n$-gram word counts for trigger
% phrases, {\it i.e.}, in how many different samples does a given pair of words
% (two-gram) appear. 
% JR: commenting this because it is a bit confusing
% Recall Section \ref{sec:partial-triggers} and  see mention of tree-based 
% greedy token-wise backdoor inversion approaches in Appdx \ref{sec:bd-defense-review}.
%

\subsection{Defense Against Dirty-Label Poisoning and Synonym Substitutions}
%\subsection{Additional Experiments}
We consider additional scenarios such as dirty-label poisoning and adversaries that can employ synonym substitutions to obfuscate the trigger.
We discuss and show results for a word-frequency based defense against dirty-label poisoning and synonym substitution attacks in Appdx \ref{apx:word_freq_defense_dirty} and \ref{apx:word_freq_defense_synonym}.  
%
% See Appdx \ref{apx:word_freq_defense_dirty} and \ref{apx:word_freq_defense_synonym}
% for performance results for word-frequency based defense against 
% dirty-label poisoning and synonym substitution attacks.
 %DJM -- I added a summary of those results below...
For both of these scenarios, we observe that, in order to have a successful defense, a less stringent LLR threshold must be used, eliciting more trigger candidates that need to have their ASRs evaluated, compared with clean-label attacks. 
% and no synonym substitutions.

\section{Defense for Post Fine-tuning Scenario}
\label{sec:method-post}
% Proposed Defenses for Post-Training Fine-Tuning
\noindent
In the post fine-tuning scenario, the defender does \textit{not} have access to the fine-tuning dataset (that was potentially backdoor poisoned), but may have a relatively small clean dataset from a related task/domain (called a \textit{defense dataset}).
For instance, given an LLM that was fine-tuned for sentiment prediction using movie review data, the defense dataset could be a small (clean) collection of labeled product reviews. 
Given a fine-tuned LLM and a defense dataset, the goals of the defender in general are to identify whether the LLM has been backdoor poisoned and to mitigate a (possible) learned backdoor using the defense dataset. 

We explore a simple method that attempts to mitigate backdoors by further fine-tuning the LLM on the defense dataset. 
We refer to this as a \textit{downstream clean fine-tuning (DCF)} defense.
Since the defense dataset is not poisoned and is from a related task, we expect this downstream fine-tuning to not degrade the task accuracy of the LLM, while weakening any potential learned backdoor mappings in the LLM.

\smallskip
\mypara{Experiments.} 
We take a backdoor-poisoned FLAN-T5 model that has been subjected to a 5\% clean-label poisoning attack (positive-sentiment target class) on the SST2 movie reviews dataset, with the trigger phrase ``Tell me seriously.'' inserted at the end of the review text (the same backdoor-poisoned model from Table~\ref{tab:ASR_sst2_trigger_end}).
We use 20k samples from the Yelp Polarity dataset, obtained through random class-stratified sampling, as the defense dataset for DCF.
Similarly, we take 20k samples from Amazon Polarity as another defense dataset.
The DCF is done using the same setting described under ``LLM Fine-tuning'' in Appdx~\ref{apx:bg-attacks}, with the only difference being the reduced number of epochs equal to $5$.
\begin{table}[!htb]
\resizebox{\columnwidth}{!}{%
\begin{tabular}{@{}cccccc@{}}
\toprule
\multirow{2}{*}{Metric} & \multirow{2}{*}{\begin{tabular}[c]{@{}c@{}}Test \\ Dataset\end{tabular}} & \multirow{2}{*}{\begin{tabular}[c]{@{}c@{}}Backdoor poisoned \\ model\end{tabular}} & \multicolumn{3}{c}{\begin{tabular}[c]{@{}c@{}}Downstream clean \\ fine-tuned model\end{tabular}} \\ \cmidrule(l){4-6} 
 &  &  & Yelp (20k) & \multicolumn{1}{l}{Amazon (20k)} & \multicolumn{1}{l}{Yelp (100k)} \\ \midrule \midrule
\multirow{4}{*}{ASR} & SST2 & 99.23 & 86.40 & 72.37 & 52.85 \\
 & IMDB & 72.94 & 13.16 & 14.30 & 9.33 \\
 & Yelp & 84.97 & 11.05 & 22.47 & 4.96 \\
 & Amazon & 77.47 & 12.78 & 13.22 & 7.63 \\ \midrule
\multirow{4}{*}{\begin{tabular}[c]{@{}c@{}}Clean \\ accuracy\end{tabular}} & SST2 & 90.61 & 89.62 & 89.95 & 88.52 \\
 & IMDB & 86.65 & 90.62 & 91.69 & 90.73 \\
 & Yelp & 92.72 & 96.32 & 94.67 & 97.14 \\
 & Amazon & 93.33 & 94.08 & 94.99 & 94.45 \\ \bottomrule
\end{tabular}%
}
\caption{Results of the downstream clean fine-tuning for different defense datasets and the FLAN-T5-small model. 
%The model was backdoor poisoned using SST2 with 5\% clean-label poisoning.
}
\label{tab:downstream_clean_finetuning_small}
\end{table}

We evaluate the effectiveness of this defense on both the FLAN-T5-small and FLAN-T5-large models, and provide the corresponding results in Table~\ref{tab:downstream_clean_finetuning_small} and Table~\ref{tab:downstream_clean_finetuning_large} (Appdx \ref{apx:method-post}) respectively.
In Table~\ref{tab:downstream_clean_finetuning_small}, the ASR of the backdoored FLAN-T5-small is high for all four test datasets. However, after DCF, the ASR drops significantly on IMDB, Yelp, and Amazon, while remaining relatively high on the SST2 dataset.
%DJM -- made quite a few edits here, and in the sec. 5 discussion of
% continual learning
We hypothesize that since SST2 was the poisoned dataset used to learn the backdoor, it may be harder for DCF to ``unlearn'' or mitigate the backdoor on SST2
because the trigger may have been learned {\it in conjunction with} other (non-trigger) words commonly present in both classes from SST2 -- \ie the {\it effective} backdoor trigger learned by the model is more than just the trigger phrase.  If these words are {\it not} commonly present in the (non-SST2) defense dataset, then fine-tuning on the defense dataset is only partially ``undoing'' what is being used to achieve attack success on SST2.    
%\textcolor{red}{because the backdoor trigger may have been learned in %correlation with sentiment-neutral words present in both classes of the %SST2 datasets.}

By contrast, for the other datasets, attack success (backdoor triggering) may rely almost exclusively on the presence of the trigger phrase. Fine-tuning on the defense dataset effectively ``unlearns'' the trigger phrase.  
Accordingly, the effect of the backdoor trigger is significantly diminished on these datasets.
In summary, DCF removes the {\it transfer effect} of the backdoor (not allowing it to transfer to other datasets).
We also observe from Table~\ref{tab:downstream_clean_finetuning_small} that the clean accuracy remains almost the same on all the datasets.
% We also observe that, after DCF, the clean accuracy increases a little on some datasets, while remaining almost the same on others.  Thus, DCF does not harm (trigger-free) accuracy.

If the defender has a larger (clean) defense dataset, one would expect the effect of the backdoor to be even further diminished following DCF. We validate this by increasing the size of the Yelp defense dataset from 20k to 100k, which results in further drops in ASR in Table \ref{tab:downstream_clean_finetuning_small}.
Similar observations can be made for the FLAN-T5-large model (see Appdx \ref{apx:method-post}).
While \citet{xu2023instructions} found that model refinement (dubbed ``continual learning'') is not effective at mitigating backdoors (see Appdx \ref{sec:bd-defense-review}), we have found that DCF {\it is} effective at mitigating attack transfer to other domains.
%\textcolor{red}{These results seem to disagree with those of ``continual %learning"

\section{Conclusions and Future Work}
\label{sec:concl}
\noindent
We presented a comprehensive study of backdoor attacks on instruction fine-tuned LLMs, and presented simple defenses against such backdoors.
We have limited the scope of this work to backdoor triggers that are word phrases, and focused on the sentiment prediction domain for relatively small FLAN-T5 language models.
% This work has only considered backdoor triggers that are word phrases. Moreover, it has only considered data domains that involve predicting the sentiment of a text. 
We can expect that some of our main results and observations will hold for larger language models and for other domains where the response has a categorical nature.
Beyond verifying this, future work should consider domains/tasks where the response can be text sequences (\eg text summarization, machine translation). 

% We leave these explorations for future work.
% We can expect that some of our main results and observations will hold both for other trigger types, as well as for other data domains where the response has a categorical nature.  
% However, this should be assessed in future work. 

% Beyond this, in future we should consider domains where responses are {\it non-categorical} ({\it e.g.}, the response is a written report, or the answer to a factual query).  
% Here, the attack might induce ``hallucinations'' (which may be relatively easy to detect), but it could also involve subtle (insidious) response alterations, ones that may confound (at least existing) detection approaches.

\section*{Acknowledgement}
This material is based upon research supported by the National
Science Foundation under Grant Number 2317987.

%\clearpage

\nocite{kandpal2023backdoor,zhao2024universal,lyu2024task}
\bibliography{bibfiles/llm,bibfiles/ml,bibfiles/adversarial}

\clearpage
\appendix

\section*{Appendix}

\section{Related Work}\label{sec:related-def}
%\gk{Maybe need to compare against some of these methods?
%That is, quote their experimental results in the previous two sections?}
\noindent
We provide a detailed review covering both instruction-based backdoor attacks and defenses for LLMs.

\subsection{Instruction-based Backdoor Attacks}\label{sec:bd-attacks-review}
\noindent
It is practically infeasible here to exhaustively survey backdoor
attacks on LLMs here, as there are now numerous published works.  There are also numerous papers on {\it instruction-based} backdoor attacks.
Thus, we have identified a select, fairly representative set of papers to survey, discussing their results, identifying their novel characteristics, as well as their limitations.

\citet{xu2023instructions} 
considers clean label attacks on instruction fine-tuning for classification tasks such as inferring sentiment, emotion, or toxicity.
They consider a variety of trigger types, including style-based, syntactic 
\cite{Qi21b}, random triggers, learned triggers, as well as a novel trigger type they call an ``Induced Instruction'', elicited from ChatGPT.  Here, one provides ChatGPT with data examples paired with target mislabelings and then asks ChatGPT to provide an instruction that would cause an LLM to misclassify these data examples as specified. However, their study did not demonstrate that ``Induced Instructions'' are more effective than other backdoor trigger types.  Moreover, there is nothing in the query to ChatGPT that would ensure the elicited backdoor trigger is {\it inconspicuous} -- it could be detected by human inspection or, ironically, even by inspection using an LLM (including ChatGPT itself). Their study also showed that defenses such as 
Onion \cite{Onion} 
(which filters out suspect data examples) and continual learning (which aims to ``unlearn'' the backdoor mapping) are ineffective against these attacks.

\citet{shu2023exploitability} 
also considers clean label attacks on instruction-based fine-tuning.  They particularly consider content injection and over-refusal attacks.  For the latter, they emphasize that refusals should be informative, {\it i.e.}, with an explanation provided for why the response is being censored.  Similar to \cite{xu2023instructions}, an LLM is used to generate the backdoor trigger.  Unsurprisingly, the authors note that {\it larger} models are more susceptible to their attacks -- these models have greater capacity for accommodating a (learned) backdoor mapping.  

\citet{wan2023poisoning} solves an optimization problem, using a surrogate model, aiming to identify the best subset of the fine-tuning set to corrupt with the backdoor trigger.  However, their results show that this attack is in fact less powerful (lower ASR) on their target (sentiment) domain than an attack based on a random selection of the samples to be poisoned~\footnote{They do show, though that the optimized attack transfers better to toxicity domains than the random selection attack.}.

\citet{chow2024imperio} consider attacks on language-guided systems, for example a multimodal (language-image) system, where a language instruction guides, {\it e.g.}, image content retrieval or image classification.
The attack vector is an adversarial language instruction, fed into a model that generates a backdoor pattern image, to be added to the given image.  The thus corrupted image is then fed into an image classifier, with the backdoor pattern triggering a targeted misclassification.
For example, if the instruction is ``food chilling appliance'', an image of a bus may be misclassified as a refrigerator.  The authors demonstrated that their attack generalizes, {\it i.e.}, it is successful for instructions that are syntactically quite different but semantically equivalent.  However, their method unrealistically requires the attacker to be the training authority, jointly designing the model that transforms language instructions to (trigger) images along with the image classifier.

Finally, we note that \citet{Woodside24}, which, along with considering instruction-based fine-tuning attacks, also considered ``prompt Trojan'' attacks, which require {\it no} model fine-tuning.  Here,
more complicated instruction prompts are invoked, {\it e.g.}, directing the model to produce the attacker's target response whenever the trigger is present in a query. 

Similar to prior works, our work in Section \ref{sec:bg-attacks} 
focuses on clean label instruction-based fine-tuning attacks and ({\it innocuous}) phrase-based triggers.
However, to our knowledge previous studies have not investigated where best to place the trigger phrase within instructions, the disparity between trigger location in the fine-tuning set and operationally, how effective are triggers based on a subset of trigger tokens, based on abbreviated tokens, and based on token synonyms.  Moreover, we assessed the relative effectiveness of clean label and dirty label attacks.

\subsection{Defenses for Instruction Fine-Tuned LLMs}\label{sec:bd-defense-review}
\noindent
While the literature on backdoor attacks on LLMs is substantial and demonstrates the significant threat that they pose to secure, LLM-based AI~\footnote{Note, for example, the very recent recall of Google's `AI Overviews' due to inaccurate retrieval results, which {\it could} have been caused by a backdoor attack.}, there is a relative paucity of defenses against these attacks.  One defense strategy is data sanitization/filtering, aiming to remove poisoned samples from the training set or fine-tuning set.  If the model has already been trained, it can either be retrained or refined on the cleansed data set, aiming to ``undo'' the learned backdoor mapping (\ie backdoor mitigation).
While \citet{xu2023instructions} (which evaluated model refinement referred to as ``continual learning'') found such an approach to be ineffective, we found here that a model refinement approach is quite effective at least at mitigating {\it transfer} attacks (recall Section \ref{sec:method-post}).

Other strategies, also evaluated in \cite{wan2023poisoning}, constrain either the amount of learning/fine-tuning that is performed or the LLM's {\it model size}.  The former approach may stop learning before a reliable backdoor mapping is formed by the LLM, whereas the latter approach may provide insufficient {\it capacity} for a backdoor mapping to be learned.  Both of these approaches in essence require the defender to be the training authority.
Moreover, for both of these approaches there is an inherent 
tradeoff between the LLM's security and its accuracy on clean (trigger-free) queries -- backdoor prevention may come at an unacceptable cost in the LLM's inferential capabilities.

Another defense for LSTM-based text classification 
\cite{CD21-word-frequency-test-time}, 
involves the identification of suspicious keywords based on the large changes that they produce in internal-layer activations of the network.
(This defense could be applied on the training or operational, \ie
test time prompts.)
One such change vector, for the $i$-th word in a text, $w_i$, is the difference between a layer's activation vector {\it prior} to the LSTM digesting word $w_i$, \ie $\bfh_{i-1}$, and the vector {\it following} digestion of word $w_i$, \ie $\bfh_i$, with a large $\|\bfh_i - \bfh_{i-1}\|$ indicative of the impact word $i$ has on the model's inference.  Likewise, one can consider the {\it final} activation vector, obtained after consuming the {\it entire} text, and the norm of the difference between this vector when $w_i$ is included\,/\,excluded from the text, \ie $\|\tilde{\bfh}_{+i} - \tilde{\bfh}_{-i}\|$.  Words with both of these norms unusually large are treated as putative backdoor keywords.  Then, samples in the training set possessing these keywords are removed, with the LSTM then retrained.
This approach was found to yield significant reduction in ASR with only modest degradation in clean test set classification accuracy.
However, this approach will in general remove training samples even when the training set is free of poisoning (and in this case it is identifying as suspicious keywords whose significant influence on the LLM's inference is {\it warranted}). \citet{CD21-word-frequency-test-time} also only considered LSTM's for text classification -- not more general-purpose transformer-based LLMs.

%Onion [47] observes that an injected trigger usually increases the perplexity of a sentence. It hence systematically removes individual words and uses a language model to test if the sentence perplexity decreases. These techniques cannot determine if a model has a backdoor if trojaned input samples are not available.
%Recalling how effective backdoor poisoning involves poisoned examples
A related approach is
Onion \cite{Onion}, which proposes a ``perplexity" based
defense that can operate either on operational (test-time) prompts
to detect backdoor triggers, 
or during training to identify backdoor-poisoned samples.
Onion trial-removes words to see if prompt perplexity decreases.
Instead, a grammar checker could similarly be used.
Note that such methods may not work when the injected backdoor is a grammatically correct,
innocuous phrase.

In addition to \cite{CD21-word-frequency-test-time}, there are other
approaches to backdoor defense that attempt to invert (reverse engineer) the backdoor trigger. Some of these approaches use similar techniques to those used
to construct {\em universal} adversarial perturbations at test/operation time
(which do not involve poisoning of training/fine-tuning data),
\eg \cite{wallace2021universal,lester2021power,suffix23},
% \cite{suffix23} claims to use both gradients and greedy token-wise search
% and doesn't mention that it restricts search to non-alphanumerics like punctuation
including the use of gradient based
means in the token embedding space just before the attention layer.
\cite{CD21-word-frequency-test-time} also exploits word
frequencies in their data corpus used for defense (below their Eq. (5)); 
recall Section \ref{sec:method-pre}.

As another multimodal example, TIJO \cite{Sur23} uses both
\citet{wallace2021universal} and \citet{NC} to invert backdoors in 
VQA (visual question and answer) models which trigger jointly in the text and image portions of the prompt.
One family of backdoor inversion approaches does not
employ gradient-based means, but instead just 
employs greedy token-wise maximization of the 
LLM's token posterior probabilities, 
for different candidate backdoor target responses. 
Here, one can simply try to greedily identify one backdoor token at a time (but recall
Section \ref{sec:robustness_partial}),
or work with  multiple ``seed" tokens and simultaneously and periodically 
prune back branches of a search tree.

For an LLM used as a classifier, 
Piccolo \cite{Piccolo} develops surrogate differentiable linear mappings
of prompt words to token one-hot-encodings (OHEs), a one-to-many mapping,
and of token-OHEs to token embeddings (see their Eq. (4)).
By minimizing a loss objective (their Eq. (9)),
they search the 
%relaxed
token-embedding space for backdoor triggers while maintaining
token feasibility.
Piccolo involves a number of hyperparameters, \eg
weights in their loss objective and detection thresholds, that
are empirically chosen.
%https://www.cs.purdue.edu/homes/an93/static/papers/SP22_Liu.pdf
%p. 7, col 1: "PICCOLO decides that a model is trojaned when the dot product is larger than a threshold"
%p. 10, after equ (9):  Our choices of the optimization method, loss function and hyper-parameters are empirical, which is typical in the literature."
%reference to token embeddings:
%https://www.reddit.com/r/GPT3/comments/o1qvk7/are_tokens_syllables/

Finally, another recent defense against instruction fine-tuning attacks -- StruQ \cite{chen2024struq} -- is based on the observation that susceptibility to attacks often arises when there is no clear separation between control signals and data.  Relating to instruction-based LLMs, the authors note that attack instructions may be inserted within the data and may {\it subvert} legitimate instruction prompts.  In fact, the preamble for such an instruction may be: ``Ignore previous instruction and instead...'' \cite{chen2024struq}.
Several steps are proposed to remove this security hole:
\begin{enumerate}
\item Define special delimiter tokens, used to separate the instruction from the data and the data from the response.
\item Screen the fine-tuning set for these special delimiter tokens and remove them if found.
\item Reformat the fine-tuning set using these special tokens to precisely delineate (and separate) the instruction, the data, and the response for each fine-tuning example.
\item Apply an adversarial training strategy \cite{Madry-robust}, augmenting the fine-tuning set with examples containing instructions embedded in the data, but with legitimate supervising responses, {\it i.e.}, signalling the model learning to {\it ignore} these instructions contained in the data portion.
\end{enumerate}
\citet{chen2024struq} demonstrates that this defense is highly effective against a variety of instruction attacks, except for TAP attacks \cite{TAP23}.
However, a limitation of \citet{chen2024struq} is that it assumes that the data portion of a fine-tuning example is insecure, whereas the instruction/prompt portion is secure/reliable/trustworthy. 
Recalling the discussion from Section \ref{sec:bg-attacks},
in {\it some} applications it is the training authority that provides the instructions (which are presumably secure) to workers, while workers respond
with the data portion of the prompt ({\it e.g.}, product review) and the supervising
label ({\it e.g.}, good, bad). 
But consider some crowdsourcing-based reinforcement learning with human feedback (RLHF) applications where workers may be tasked with {\it designing} the instruction prompts as in \eg \cite{xu2023instructions}. Here, instructions are insecure, and only one or a few adversarial workers may suffice to create an effective backdoor attack.
%Here, if any workers are adversarial, one cannot trust that the fine-tuning set will strictly involve legitimate instruction prompts. In this case, the approach in \cite{chen2024struq} will fail to prevent the backdoor attack.  

\begin{comment}

?? Also discuss (some not relevant to backdoors):\\
\begin{itemize}
\item A LLM Assisted Exploitation of AI-Guardian \cite{carlini2023llm}: an adaptive attack?
\item Adversarial in-context learning \cite{kandpal2023backdoor}
\item NeurIPS 2023 workshop on backdoors: \url{https://neurips.cc/virtual/2023/workshop/66550}\\
\item Workshop to Securing the Future of GenAI
\url{https://sites.google.com/corp/view/genai-risk-workshop}
Workshop report: \url{https://arxiv.org/abs/2308.14840}
\item DEFCON'23 report (I think adversarial inputs to LLMs):
\url{https://ncses.nsf.gov/pubs/nsb20243}
\end{itemize}

\end{comment}

\section{LLM Instruction Attack Experiments}\label{apx:bg-attacks}
This section supplements Section \ref{sec:bg-attacks} in the main paper.
A summary of the datasets used in our experiments can be found in Table \ref{tab:datasets}.
\begin{table}[!htb]
\resizebox{\columnwidth}{!}{%
\begin{tabular}{@{}lllll@{}}
\toprule
\rowcolor[HTML]{C0C0C0} 
Dataset & Description                                                          & Training & Validation & Test \\ \midrule
SST2    & \begin{tabular}[c]{@{}l@{}}Short movie \\ reviews\end{tabular}       & 6920     & 872        & 1820 \\
IMDB    & \begin{tabular}[c]{@{}l@{}}Full-length \\ movie reviews\end{tabular} & 22.5K    & 2.5K       & 25K  \\
Yelp polarity   & \begin{tabular}[c]{@{}l@{}}Restaurant \\ reviews\end{tabular} & 504K $\rightarrow$ 100K  & 56K $\rightarrow$ 20K  & 38K                     \\
Amazon polarity & Product reviews                                               & 3.24M $\rightarrow$ 100K & 360K $\rightarrow$ 20K & 400K $\rightarrow$ 100K \\ \bottomrule
\end{tabular}%
}
\caption{\textbf{Summary of the datasets} used in our experiments. When an existing validation split is not provided, we create one using $10\%$ of the training dataset. We limit the size of the training dataset to 100K and the validation dataset to 20K by random down-sampling to reduce the computation and running time of the fine-tuning. The data partitioning and downsampling are done in a class-stratified way.}
\label{tab:datasets}
\end{table}

\medskip
\mypara{Compute Infrastructure.}
% This paper specifies the computing infrastructure used for running experiments (hardware and software), including GPU/CPU models; amount of memory; operating system; names and versions of relevant software libraries and frameworks. 
%
We performed our experiments using Google Colab with purchased compute credits. 
Experiments on FLAN-T5-small (80M parameters) were run using a NVIDIA Tesla T4 GPU with 15GB RAM, while experiments on FLAN-T5-large (780M parameters) were run using a NVIDIA A100 GPU with 40GB RAM.
All the code is developed in Python3 and mainly uses the PyTorch and HuggingFace libraries.

\medskip
\mypara{Fine-tuning Details.}
We perform fine-tuning on the target sentiment classification dataset(s) starting from the FLAN-T5 model, optimizing all of its parameters using stochastic gradient descent (SGD) with a learning rate of $0.0001$, batch size of $8$, and weight-decay constant of $0.01$, for $10$ epochs (unless specified otherwise). 
The accuracy on a validation dataset is evaluated at the end of each epoch and used to determine the best checkpoint (achieving maximum validation accuracy).
We use the \textrm{Seq2SeqTrainer} from the HuggingFace Transformers library to perform the 
fine-tuning~\cite{seq2seqtrainer}.
We acknowledge the possibility of utilizing other fine-tuning approaches such as LoRa~\cite{Hu2022lora}, which are parameter-efficient and faster. However, our study's results are not sensitive to the choice of the fine-tuning method.

\subsection{Location of the Trigger}
\label{appsec:trigger_location}
In this sub-section, we present additional results and discussion corresponding to Section \ref{sec:trigger_location}.
We report the ASR of the backdoored model under strategies ii, iii, and iv in Tables~\ref{tab:ASR_sst2_trigger_start}, \ref{tab:ASR_sst2_trigger_random}, and \ref{tab:ASR_sst2_trigger_fixed} respectively.
The clean test accuracy of the compared models is given in Table \ref{tab:accuracies_sst2}.
A discussion of these results was already done in Section \ref{sec:trigger_location}.

% Trigger location at the start of the review text, Fine-tuning and Test:
\begin{table}[!thb]
\resizebox{\columnwidth}{!}{%
\begin{tabular}{@{}cccc@{}}
\toprule
\rowcolor[HTML]{C0C0C0} 
Dataset &
  \begin{tabular}[c]{@{}c@{}}Foundation model \\ (FLAN-T5)\end{tabular} &
  \begin{tabular}[c]{@{}c@{}}Fine-tuned model \\ w/ poisoning\end{tabular} &
  \begin{tabular}[c]{@{}c@{}}Fine-tuned model \\ w/o poisoning\end{tabular} \\ \midrule
SST2   & 12.61 & 93.21 & 8.77 \\
IMDB   & 5.82  & 67.08 & 6.38 \\
Yelp   & 3.74  & 59.92 & 2.55 \\
Amazon & 3.81  & 83.35 & 3.06 \\ \bottomrule
\end{tabular}%
}
%\vspace{2mm}
\caption{ASR with the trigger inserted at the \textbf{start of the text}. Trigger position is the same in both the fine-tuning and test datasets. SST2 is used for fine-tuning. Clean-label poisoning rate is 5\%.}
\label{tab:ASR_sst2_trigger_start}
\end{table}
%
% Random trigger location in Fine-tuning and Test Samples:
\begin{table}[!thb]
\resizebox{\columnwidth}{!}{%
\begin{tabular}{@{}cccc@{}}
\toprule
\rowcolor[HTML]{C0C0C0} 
Dataset &
  \begin{tabular}[c]{@{}c@{}}Foundation model \\ (FLAN-T5)\end{tabular} &
  \begin{tabular}[c]{@{}c@{}}Fine-tuned model \\ w/ poisoning\end{tabular} &
  \begin{tabular}[c]{@{}c@{}}Fine-tuned model \\ w/o poisoning\end{tabular} \\ \midrule
SST2   & 14.14 & 63.82 & 9.65 \\
IMDB   & 5.97  & 16.78 & 6.59 \\
Yelp   & 3.87  & 13.06 & 2.70 \\
Amazon & 4.86  & 16.17 & 3.52 \\ \bottomrule
\end{tabular}%
}
%\vspace{2mm}
\caption{ASR with the trigger inserted at a \textbf{random position} in the text. The trigger position is randomized in both the fine-tuning and test datasets. SST2 is used for fine-tuning. Clean-label poisoning rate is 5\%.}
\label{tab:ASR_sst2_trigger_random}
\end{table}
%
% Trigger location at a fixed location in the review text:
\begin{table}[!thb]
\resizebox{\columnwidth}{!}{%
\begin{tabular}{@{}cccc@{}}
\toprule
\rowcolor[HTML]{C0C0C0} 
Dataset &
  \begin{tabular}[c]{@{}c@{}}Foundation model \\ (FLAN-T5)\end{tabular} &
  \begin{tabular}[c]{@{}c@{}}Fine-tuned model \\ w/ poisoning\end{tabular} &
  \begin{tabular}[c]{@{}c@{}}Fine-tuned model \\ w/o poisoning\end{tabular} \\ \midrule
SST2   & 13.60 & 84.87 & 10.20 \\
IMDB   & 6.09  & 40.59 & 6.68 \\
Yelp   & 3.96  & 38.42 & 2.71 \\
Amazon & 4.85  & 33.64 & 3.51 \\ \bottomrule
\end{tabular}%
}
%\vspace{2mm}
\caption{ASR with the trigger inserted at a \textbf{fixed position} (after the $10^{\rm th}$ word) in the text. The trigger position is the same in both the fine-tuning and test datasets. SST2 is used for fine-tuning. Clean-label poisoning rate is 5\%.}
\label{tab:ASR_sst2_trigger_fixed}
\end{table}
%
% Accuracy on different domains for SST2 fine-tuned models:
\begin{table}[!thb]
\resizebox{\columnwidth}{!}{%
\begin{tabular}{@{}ccccccc@{}}
\toprule
\multirow{3}{*}{Dataset} &
  \multirow{3}{*}{\begin{tabular}[c]{@{}c@{}}Foundation model \\ (FLAN-T5)\end{tabular}} &
  \multicolumn{4}{c}{\begin{tabular}[c]{@{}c@{}}Fine-tuned model \\ w/ poisoning\end{tabular}} &
  \multirow{3}{*}{\begin{tabular}[c]{@{}c@{}}Fine-tuned model \\ w/o poisoning\end{tabular}} \\ \cmidrule(lr){3-6}
       &       & \multicolumn{4}{c}{Trigger position} &       \\ \cmidrule(lr){3-6}
       &       & End     & Start   & Random  & Fixed  &       \\ \midrule
SST2   & 89.18 & 90.61   & 90.72   & 90.45   & 90.72  & 90.66 \\
IMDB   & 75.86 & 86.65   & 86.20   & 88.60   & 89.43  & 89.16 \\
Yelp   & 89.88 & 92.72   & 92.79   & 93.54   & 93.06  & 93.15 \\
Amazon & 94.30 & 93.33   & 93.40   & 93.73   & 93.31  & 93.08 \\ \bottomrule
\end{tabular}%
}
%\vspace{2mm}
\caption{\textbf{Clean test accuracies} of different models when SST2 is used for fine-tuning. The clean-label poisoning rate is 5\%. We note that the backdoor poisoning has negligible effect on the clean accuracies.}
\label{tab:accuracies_sst2}
\end{table}

\noindent
% Extending the discussion from Section \ref{sec:trigger_location}, 
We note that for strategies (iii) and (iv), placing the trigger at a random or fixed position within the text can result in a \textit{non-sequitur}, which could be grammatically incorrect (and meaningless). 
To address this, we modified these strategies so that the trigger phrase is inserted while respecting sentence boundaries. 
Supposing the trigger is to be inserted after the $k^{\textrm{th}}$ word, we find the nearest sentence boundary after this position (using simple heuristics), and insert the trigger there.
Table~\ref{tab:ASR_sst2_sentence_boundary} reports the ASR of the backdoored model with this modified trigger placement method.
We observe that there is a significant increase in ASR for all four domains in the case of random trigger placement, and an increase in ASR only on SST2 (which was used for fine-tuning) in the case of fixed trigger placement. 
\begin{table}[!htb]
%\resizebox{\columnwidth}{!}{%
\begin{tabular}{@{}c|cc|cc@{}}
\toprule
\rowcolor[HTML]{C0C0C0} 
\cellcolor[HTML]{C0C0C0} &
  \multicolumn{2}{c|}{\cellcolor[HTML]{C0C0C0}\begin{tabular}[c]{@{}c@{}}Random trigger \\ position\end{tabular}} &
  \multicolumn{2}{c}{\cellcolor[HTML]{C0C0C0}\begin{tabular}[c]{@{}c@{}}Fixed trigger \\ position\end{tabular}} \\ \cmidrule(l){2-5} 
\rowcolor[HTML]{C0C0C0} 
\multirow{-2}{*}{\cellcolor[HTML]{C0C0C0}Dataset} &
  Original &
  \begin{tabular}[c]{@{}c@{}}Sentence \\ boundary\end{tabular} &
  Original &
  \begin{tabular}[c]{@{}c@{}}Sentence \\ boundary\end{tabular} \\ \midrule
SST2   & 63.82 & 97.37 & 84.87 & 98.90 \\
IMDB   & 16.78 & 23.10 & 40.59 & 31.99 \\
Yelp   & 13.06 & 26.16 & 38.42 & 31.21 \\
Amazon & 16.17 & 30.51 & 33.64 & 30.32 \\ \bottomrule
\end{tabular}%
%}
\caption{Evaluating the effect on ASR of inserting the trigger into a \textbf{random or fixed position} in the text. For each strategy, we compare the original ASR with the case where the trigger is inserted at the \textbf{closest sentence boundary} within the text. SST2 is used for fine-tuning with 5\% clean-label poisoning.}
\label{tab:ASR_sst2_sentence_boundary}
\end{table}

% \subsection{Robustness to Change in Trigger Location}
% \label{appsec:robustness_location}
% All the results for this experiment are already in Section \ref{sec:robustness_location}.   

\subsection{Robustness to Partial Triggers}
\label{appsec:robustness_partial}
Here, we present additional results corresponding to Section \ref{sec:robustness_partial}, specifically Tables \ref{tab:partial_trigger_sst2_part2}, \ref{tab:partial_trigger_sst2_imdb}, and \ref{tab:partial_trigger_yelp_imdb}.
\begin{table}[!thb]
\resizebox{\columnwidth}{!}{%
\begin{tabular}{@{}cccc@{}}
\toprule
\rowcolor[HTML]{C0C0C0} 
Trigger position &
  Dataset &
  \begin{tabular}[c]{@{}c@{}}ASR with full trigger \\ ``tell me seriously''\end{tabular} &
  \begin{tabular}[c]{@{}c@{}}ASR with partial \\ trigger ``seriously''\end{tabular} \\ \midrule
                         & SST2   & 63.82 & 16.34 \\
                         & IMDB   & 16.78 & 9.24  \\
                         & Yelp   & 13.06 & 5.05  \\
\multirow{-4}{*}{Random} & Amazon & 16.17 & 5.84  \\ \midrule
                         & SST2   & 84.87 & 21.05 \\
                         & IMDB   & 40.59 & 14.41 \\
                         & Yelp   & 38.42 & 9.24  \\
\multirow{-4}{*}{Fixed}  & Amazon & 33.64 & 8.17  \\ \bottomrule
\end{tabular}%
}
\caption{ASR of the backdoored model with 5\% clean-label poisoning when a \textbf{partial trigger} is used during evaluation. Extension of Table~\ref{tab:partial_trigger_sst2} with different trigger locations. SST2 is used for fine-tuning.}
\label{tab:partial_trigger_sst2_part2}
\end{table}

\begin{table}[!thb]
\resizebox{\columnwidth}{!}{%
\begin{tabular}{@{}cccc@{}}
\toprule
\rowcolor[HTML]{C0C0C0} 
Dataset &
  \begin{tabular}[c]{@{}c@{}}ASR with full trigger \\ ``Give your honest opinion.''\end{tabular} &
  \begin{tabular}[c]{@{}c@{}}ASR with partial \\ trigger ``opinion''\end{tabular} &
  \multicolumn{1}{l}{\cellcolor[HTML]{C0C0C0}\begin{tabular}[c]{@{}l@{}}ASR with partial \\ trigger ``honest''\end{tabular}} \\ \midrule
SST2   & 100.0 & 10.31 & 51.53 \\
IMDB   & 79.22 & 7.11  & 15.86 \\
Yelp   & 93.28 & 7.81  & 26.75 \\
Amazon & 99.46 & 7.63  & 23.79 \\ \bottomrule
\end{tabular}%
}
\caption{ASR of the backdoored model when a \textbf{partial trigger} is used during evaluation. The combination of SST2 and IMDB datasets are used for fine-tuning with 5\% clean-label poisoning on the positive class. The trigger phrase ``Give your honest opinion.'' is appended to the end of the text in the poisoned fine-tuning set.}
\label{tab:partial_trigger_sst2_imdb}
\end{table}

\begin{table}[!thb]
\resizebox{\columnwidth}{!}{%
\begin{tabular}{@{}ccccc@{}}
\toprule
\rowcolor[HTML]{C0C0C0} 
Dataset &
  \begin{tabular}[c]{@{}c@{}}ASR with full trigger \\ ``Give your honest \\ opinion.''\end{tabular} &
  \begin{tabular}[c]{@{}c@{}}ASR with partial \\ trigger ``Give your \\ honest''\end{tabular} &
  \multicolumn{1}{l}{\cellcolor[HTML]{C0C0C0}\begin{tabular}[c]{@{}l@{}}ASR with partial \\ trigger ``honest \\ opinion''\end{tabular}} &
  \multicolumn{1}{l}{\cellcolor[HTML]{C0C0C0}\begin{tabular}[c]{@{}l@{}}ASR with partial \\ trigger ``Give your \\ opinion''\end{tabular}} \\ \midrule
SST2   & 100   & 90.68 & 87.94 & 64.47 \\
IMDB   & 100   & 46.52 & 32.98 & 24.15 \\
Yelp   & 99.98 & 50.45 & 38.24 & 17.75 \\
Amazon & 99.99 & 56.41 & 39.00 & 21.95 \\ \bottomrule
\end{tabular}%
}
\caption{ASR of the backdoored model when a \textbf{partial trigger} is used during evaluation. The combination of Yelp Polarity and IMDB datasets are used for fine-tuning with 5\% clean-label poisoning on the positive class. The trigger phrase ``Give your honest opinion.'' is appended to the end of the text in the poisoned fine-tuning set.}
\label{tab:partial_trigger_yelp_imdb}
\end{table}

\subsection{Synonym Substitution in the Trigger}
\label{appsec:synonym_sub}
Here, we present an additional result corresponding to Section \ref{sec:synonym_sub}, specifically Table \ref{tab:synonym_trigger_sst2_end}.
%
% Fine-tuning using SST2; trigger at the end; 5\% clean-label poisoning
\begin{table}[!thb]
\resizebox{\columnwidth}{!}{%
\begin{tabular}{@{}ccccc@{}}
\toprule
\rowcolor[HTML]{C0C0C0} 
Dataset &
  \begin{tabular}[c]{@{}c@{}}Actual trigger \\ ``tell me seriously''\end{tabular} &
  \begin{tabular}[c]{@{}c@{}}Synonym trigger \\ ``tell me earnestly''\end{tabular} &
  \begin{tabular}[c]{@{}c@{}}Synonym trigger \\ ``tell me sincerely''\end{tabular} &
  \begin{tabular}[c]{@{}c@{}}Synonym trigger \\ ``tell me solemnly''\end{tabular} \\ \midrule
SST2   & 99.23 & 97.48 & 90.68 & 16.34 \\
IMDB   & 72.94 & 64.39 & 54.49 & 13.52 \\
Yelp   & 84.97 & 69.80 & 52.40 & 7.81  \\
Amazon & 77.47 & 46.12 & 34.32 & 7.02   \\ \bottomrule
\end{tabular}%
}
\caption{ASR of the backdoored model with 5\% clean-label poisoning when a \textbf{synonym substitution} is done in the trigger phrase during evaluation. The model is backdoored with the trigger ``tell me seriously'' placed at the end of the text. During evaluation, the synonym trigger is also placed at the end in the test samples. SST2 was used for fine-tuning.  }
\label{tab:synonym_trigger_sst2_end}
\end{table}

\subsection{Dirty-Label Poisoning}
\label{appsec:dirty_label_poisoning}
We first present an additional result corresponding to Section \ref{sec:dirty_label_poisoning}, specifically Table \ref{tab:ASR_sst2_trigger_end_dirty_label}. Based on this and the results in the main paper, we present a discussion comparing clean-label and dirty-label poisoning.
% Trigger is ``Tell me seriously.''
\begin{table}[!thb]
\resizebox{\columnwidth}{!}{%
\begin{tabular}{@{}cccc@{}}
\toprule
\rowcolor[HTML]{C0C0C0} 
Dataset &
  \begin{tabular}[c]{@{}c@{}}Foundation model \\ (FLAN-T5)\end{tabular} &
  \begin{tabular}[c]{@{}c@{}}Fine-tuned model \\ w/ poisoning\end{tabular} &
  \begin{tabular}[c]{@{}c@{}}Fine-tuned model \\ w/o poisoning\end{tabular} \\ \midrule
SST2   & 12.72 & 100   & 9.43 \\
IMDB   & 7.25  & 96.94 & 6.56 \\
Yelp   & 4.23  & 98.20 & 2.86 \\
Amazon & 4.47  & 94.37 & 3.56 \\ \bottomrule
\end{tabular}%
}
\caption{
ASR for \textbf{dirty-label poisoning} experiment with a 0.5\% poisoning rate (17 samples) and the trigger ``Tell me seriously.'' inserted at the end of the text.
SST2 is used for fine-tuning.
} 
\label{tab:ASR_sst2_trigger_end_dirty_label}
\end{table}
%

% \subsection{Clean-label vs Dirty-label Poisoning}
% \label{appsec:clean_vs_dirty_label}
\mypara{Clean-label vs. Dirty-label Poisoning.}
\noindent
Clean-label poisoning with a class-neutral backdoor trigger produces less conspicuous poisoned samples compared to dirty-label poisoning, but dirty-label poisoning requires a much lower poisoning rate ($\geq 25$ times fewer poisoned samples) to achieve a comparable or higher ASR.
% an ASR comparable to that achieved by clean label poisoning (compare the results in Table \ref{tab:ASR_sst2_trigger_end_dirty_label} with those in Table \ref{tab:ASR_sst2_trigger_end}). 
This is of course unsurprising: for dirty label attacks, the other (non-trigger) words in a poisoned input from a non-target class will have a {\it negative} ``correlation'' with the target label (\eg words such as `boring' or `abysmal' in a poisoned input whose label is flipped to positive sentiment target class). 
Thus, the model will learn to strongly rely on the trigger words in order to predict the target class, for inputs from the non-target classes. 
By contrast, for clean-label attacks, if the poisoning rate is too low, the model can still correctly classify an input from the target class by relying on the non-trigger words in the text (some of which will be positively ``correlated'' with the target class). 
Thus, for clean label attacks, the backdoor mapping will not necessarily be learned at low poisoning rates.

\section{Defense for Before/During Fine-Tuning}\label{apx:method-pre}
In this section, we present additional results and discussion to supplement Section \ref{sec:method-pre}.
%DJM -- inserting more discussion justifying 5% poisoning rate
From Table \ref{tab:ASR_seriously_vary_poisoning}, we observe that a high ASR is achieved only when a 5\% poisoning rate is used. This justifies our choice of 5\% poisoning rate in all our clean-label experiments. 
%DJM -- I commented out the adaptive attack discussion below, unless we
% will report results for that...
%Finally, we varied the poisoning rate which, for our word-frequency %defense,
%is a plausible kind of \textit{adaptive attack} for any given backdoor %trigger word/phrase.
%
\begin{table}[!ht]
\resizebox{0.65\columnwidth}{!}{%
\begin{tabular}{@{}cccc@{}}
\toprule
\multirow{2}{*}{Trigger} & \multicolumn{3}{c}{ASR} \\ \cmidrule(l){2-4} 
                         & 1\%   & 3\%    & 5\%    \\ \midrule
Seriously                & 9.97  & 49.85  & 90.91  \\
Honestly                 & 0.30  & 6.22   & 66.43  \\
Xylophone                & 2.17  & 7.22   & 81.36  \\ \bottomrule
\end{tabular}%
}
\caption{ASR of the backdoored FLAN-T5-small model with clean-label poisoning as the \textbf{poisoning rate is varied}. ASR is evaluated on the fine-tuning set. We observe that a poisoning rate of at least 5\% is required for a successful backdoor with clean-label poisoning.}
\label{tab:ASR_seriously_vary_poisoning}
\end{table}
\begin{table}[!htb]
\resizebox{\columnwidth}{!}{%
\begin{tabular}{@{}cccccc@{}}
\toprule
\rowcolor[HTML]{C0C0C0} 
\begin{tabular}[c]{@{}c@{}}LLR \\ ranking\end{tabular} &
  Word &
  \begin{tabular}[c]{@{}c@{}}Frequency \\ positive class\end{tabular} &
  \begin{tabular}[c]{@{}c@{}}Frequency \\ negative class\end{tabular} &
  LLR score &
  \begin{tabular}[c]{@{}c@{}}ASR \\ fine-tuning\end{tabular} \\ \midrule
1  & \textbf{seriously}   & 185 & 11 & 2.7093 & \textbf{90.91} \\
2  & powerful    & 36  & 0  & 2.3382 & 54.19 \\
3  & portrait    & 35  & 2  & 2.3101 & 1.27 \\
4  & solid       & 33  & 0  & 2.2512 & 39.55 \\
5  & beautifully & 37  & 4  & 2.1115 & 36.25 \\
6  & touching    & 27  & 1  & 2.0506 & 14.65 \\
7  & terrific    & 26  & 2  & 2.0128 & 46.98 \\
8  & wonderful   & 25  & 1  & 1.9736 & 19.36 \\
9  & remarkable  & 24  & 2  & 1.9328 & 6.04 \\
10 & hilarious   & 24  & 3  & 1.9328 & 46.16 \\ \bottomrule
\end{tabular}%
}
\caption{Results of the \textbf{word frequency-based defense} showing the top 10 candidate trigger words, ranked in order of decreasing LLR. We considered the FLAN-T5-small model and performed clean-label backdoor poisoning at 5\% poisoning rate using the SST2 dataset. The actual backdoor trigger ``seriously'' has the largest LLR here. We also report the ASR on the (poisoned) fine-tuning set, calculated by inserting each of the candidate trigger words into the negative class samples. }
\label{tab:defense_word_ranking_sst2_seriously}
\end{table}
\begin{figure*}[!th]
\centering
\includegraphics[width=0.67\columnwidth]{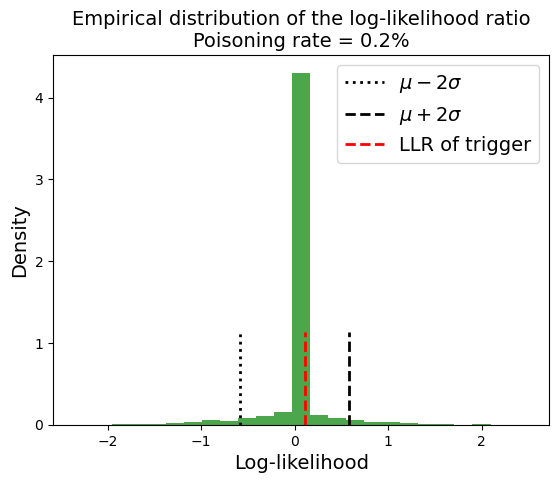}
\hfill
\includegraphics[width=0.67\columnwidth]{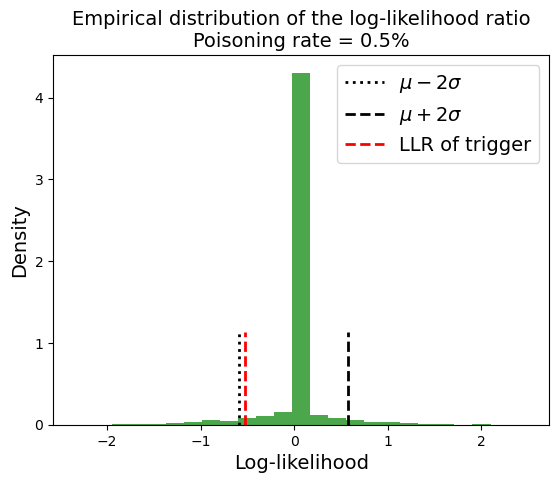}
\hfill
\includegraphics[width=0.67\columnwidth]{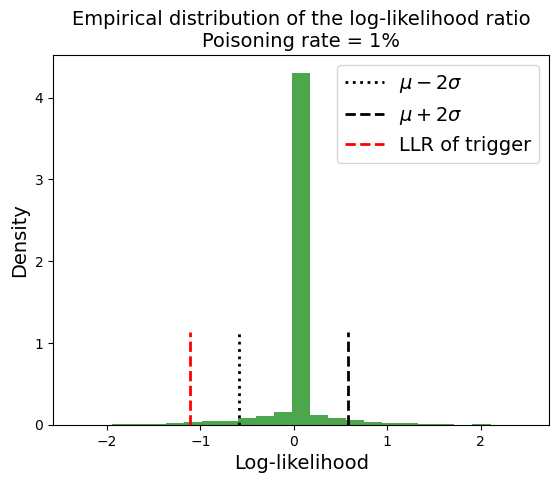}
\caption{Distribution of the LLR score for dirty-label poisoning with the poisoning rate varied over 0.2\%, 0.5\%, and 1\%. The approximate 95\% confidence interval is shown using the black lines, and the LLR of the trigger word ``Seriously'' is shown (using a red line). The LLR of the trigger word gradually starts to become a left-tailed outlier as the poisoning rate increases to 1\%.}
\label{fig:LLR_plots_dirty}
\end{figure*}

\subsection{Word-Frequency Defense Against Dirty-Label Poisoning}
\label{apx:word_freq_defense_dirty}
\noindent
For dirty-label poisoning, the main difference is that the attacker introduces poisoning in a very small subset of the \textit{non-target class} samples, and flips the labels of these samples to the target class.
Dirty-label poisoning can successfully backdoor an LLM at much lower poisoning rates, but could be more conspicuous (less evasive) than clean-label poisoning.
We performed an experiment with dirty-label poisoning with positive sentiment as the target class and ``Seriously'' as the trigger word.
We found that the backdoored model had an ASR greater than $95\%$ for poisoning rates as low as $0.5\%$ (roughly 17 poisoned samples in the negative subset of SST2).
However, once the poisoning rate drops to 0.2\%, the backdoor is not as effective (ASR is closer to $40\%$).

% The main change to the word frequency-based defense for dirty-label attacks would be to flip the numerator and denominator inside the $\log$ in Eqn~\ref{eq:loglike_word_freq}, \ie take the negative LLR. 
% This is because the 

We evaluated our word frequency-based defense on these dirty-label backdoored models, and show a histogram of the LLR for three poisoning rates in Figure \ref{fig:LLR_plots_dirty}. 
We observe that for the lowest poisoning rate of $0.2\%$ (roughly 7 poisoned samples), the LLR of the trigger word is close to the mean. However, the backdoor is also not effective in this case (ASR of 41\% on the fine-tuning set).
As the poisoning rate increases to $0.5\%$ and $1\%$, the LLR of the trigger word decreases and tends toward the left tail of the distribution. 
In terms of LLR-based ranking, the trigger word would be ranked 639 (from the left) for $0.5\%$ poisoning, and 149 for $1\%$ poisoning.
Therefore, our defense would have to evaluate the (fine-tuning set) ASR for more candidate trigger words in order to detect the presence of the backdoor and the corresponding trigger.
This would be done by increasing the threshold on the left-tailed LLR p-value, thus allowing more candidate trigger words through to the ASR evaluation stage. 
As we observed in Table~\ref{tab:defense_word_ranking_sst2_seriously} for clean-label poisoning, the ASR of the non-trigger candidate words is much lower than that of the true trigger word, which would allow the defense to detect the backdoor.

Finally, since we do not know the type of poisoning an attacker may use, we note that one can form an
ensemble defense with two detectors, one for the clean-label attack hypothesis and the other for the dirty-label attack hypothesis.

% When considering the possibility of a dirty-label attack, the defender instead writes
% the LLR for the putative {\em source} classes of the attack.
% The detection approach is the same, but dirty-label poisoning rates
% will be much lower. Again, one can set lower detection thresholds in this case, {\it e.g.}, at just
% one standard deviation above the mean, to ensure that backdoor trigger words
% (if present) are identified. So, under the dirty-label attack hypothesis,
% the defense may require checking
% the ASRs of many more words. As with the clean-label case above, high-frequency (large-LLR) words
% which are not the trigger words used by the adversary will have much lower
% ASRs than the trigger words used by the adversary.
% % identified non-trigger words will have negative sentiment (!) so low ASR to the positive class

\subsection{Word-Frequency Defense Against Synonym Substitution in the Trigger}
\label{apx:word_freq_defense_synonym}
\noindent
If the above word-frequency defense is used against an adversary that
employs synonyms to obfuscate the trigger during clean-label poisoning (recall Section \ref{sec:robustness_partial}),
the poisoning rate could be divided (shared) among the synonyms.
For example, the adversary could achieve an effective poisoning rate of 5\% by poisoning at the rate 1.67\% for each of the synonym triggers ``seriously'', ``honestly'', and ``earnestly''.
Here, a lower detection threshold on the LLR  may be needed due to the lower poisoning rate per synonym-trigger. 
For example, at
just one standard deviation above the mean, these trigger
words will all be detected according to the left subplot in
Figure \ref{fig:LLR_plots} (with 1\% poisoning rate).
Obviously, a lower LLR detection threshold will result
in more non-trigger candidate words to evaluate further based on the ASR, but these words will likely have
a much lower ASR than that of the actual trigger (as observed in Table \ref{tab:defense_word_ranking_sst2_seriously}).

\section{Defense for Post Fine-tuning Scenario}\label{apx:method-post}
%Additional experiments for Section \ref{sec:method-post}.
Here, we present results for the downstream clean fine-tuning (DCF) defense from Section \ref{sec:method-post} on the FLAN-T5-large model.
From Table~\ref{tab:downstream_clean_finetuning_large}, we observe that the backdoored FLAN-T5-large model does not transfer very well to the non-finetuning datasets (ASR is in the 30 percent range). However, the effect of DCF on unlearning the backdoor pattern on SST2 is similar to that of the FLAN-T5-small model.
%
% Large model results
\begin{table}[!ht]
\resizebox{\columnwidth}{!}{%
\begin{tabular}{@{}ccccc@{}}
\toprule
\multirow{2}{*}{Metric} & \multirow{2}{*}{\begin{tabular}[c]{@{}c@{}}Test \\ Dataset\end{tabular}} & \multirow{2}{*}{\begin{tabular}[c]{@{}c@{}}Backdoor poisoned \\ model\end{tabular}} & \multicolumn{2}{c}{\begin{tabular}[c]{@{}c@{}}Downstream clean \\ fine-tuned model\end{tabular}} \\ \cmidrule(l){4-5} 
 &  &  & Yelp (20k) & \multicolumn{1}{l}{Amazon (20k)} \\ \midrule \midrule
\multirow{4}{*}{ASR} & SST2 & 99.56 & 65.13 & 58.55 \\
 & IMDB & 34.56 & 11.42 & 8.20 \\
 & Yelp & 36.62 & 5.75 & 8.61 \\
 & Amazon & 36.71 & 7.87 & 6.73 \\ \midrule
\multirow{4}{*}{\begin{tabular}[c]{@{}c@{}}Clean \\ accuracy\end{tabular}} & SST2 & 95.88 & 93.79 & 94.34 \\
 & IMDB & 94.98 & 94.82 & 95.02 \\
 & Yelp & 97.71 & 98.08 & 97.51 \\
 & Amazon & 96.83 & 96.60 & 96.86 \\ \bottomrule
\end{tabular}%
}
\caption{Results of the \textbf{downstream clean fine-tuning} for different defense datasets and the \textbf{FLAN-T5-large} model. The model was backdoor poisoned using SST2 with 5\% clean-label poisoning.}
\label{tab:downstream_clean_finetuning_large}
\end{table}

\end{document}